\documentclass[epj,nopacs]{svjour}

\usepackage{slashed}
\usepackage{soul}
\usepackage{mathrsfs,bm}
\usepackage{txfonts}
\usepackage{amssymb}

\usepackage{amsmath}
\usepackage{indentfirst}
\usepackage{graphicx,booktabs}
\usepackage{multirow}
\usepackage{overpic}
\usepackage{color}

\usepackage{hyperref}
\usepackage{doi}
\usepackage{graphics}
\usepackage{epsfig}
\usepackage{soul}
\usepackage{color}

\usepackage[sort&compress,numbers]{natbib}

\usepackage{hyperref}

\usepackage[utf8]{inputenc}

\begin{document}
\title{The $P_{cs}(4459)$ pentaquark from a combined effective field theory
  and phenomenological perspective}
\author{Fang-Zheng Peng \inst{1} \and
  Mao-Jun Yan \inst{1} \and
  Mario {S\'anchez S\'anchez} \inst{2} \and
  Manuel {Pavon} Valderrama \inst{1,3} \mail{mpavon@buaa.edu.cn}}

\institute{School of Physics, Beihang University, Beijing 100191, China \and
Centre d'\'Etudes Nucl\'eaires, CNRS/IN2P3, Universit\'e de Bordeaux, 33175 Gradignan, France \and
International Research Center for Nuclei and Particles
  in the Cosmos \& \\
  Beijing Key Laboratory of Advanced Nuclear Materials and Physics, \\
  Beihang University, Beijing 100191, China}

\date{\today}
\abstract{
  The observation of the $P_{cs}(4459)$ by the LHCb collaboration
  adds a new member to the set of known hidden-charm pentaquarks,
  which includes the $P_c(4312)$, $P_c(4440)$ and $P_c(4457)$.
  The $P_{cs}(4459)$ is expected to have the light-quark content of
  a $\Lambda$ baryon ($I=0$, $S=-1$), but its spin is unknown.
  Its closeness to the $\bar{D}^* \Xi_c$ threshold --- $4478\,{\rm MeV}$ in
  the isospin-symmetric limit --- suggests the molecular hypothesis
  as a plausible explanation for the $P_{cs}(4459)$.
  While in the absence of coupled-channel dynamics heavy-quark spin
  symmetry predicts the two spin-states of the $\bar{D}^* \Xi_c$
  to be degenerate, power counting arguments indicate that
  the coupling with the nearby $\bar{D} \Xi_c'$ and $\bar{D} \Xi_c^*$ channels
  might be a leading order effect.
  This generates a hyperfine splitting in which the $J=\tfrac{3}{2}$
  $\bar{D}^* \Xi_c$ pentaquark will be lighter than the $J=\tfrac{1}{2}$
  configuration, which we estimate to be
  of the order of $5-15\,{\rm MeV}$.
  We also point out an accidental symmetry between the $P_{cs}(4459)$
  and $P_c(4440/4457)$ potentials.
  Finally, we argue that the spectroscopy and the $J/\psi \Lambda$ decays of
  the $P_{cs}(4459)$ might suggest a marginal preference
  for $J = \tfrac{3}{2}$ over $J = \tfrac{1}{2}$.
}


\maketitle

\section{Introduction}

The discovery by the LHCb collaboration of three hidden{-}charm
pentaquarks~\cite{Aaij:2019vzc} --- the $P_c(4312)$, $P_c(4440)$
and $P_c(4457)$ --- has triggered intense theoretical efforts to decode
their nature, in particular whether they are molecular~\cite{Chen:2019bip,Chen:2019asm,He:2019ify,Liu:2019tjn,Shimizu:2019ptd,Guo:2019kdc,Xiao:2019aya,Fernandez-Ramirez:2019koa,Wu:2019rog,Valderrama:2019chc} or not~\cite{Eides:2019tgv,Wang:2019got,Cheng:2019obk,Ferretti:2020ewe,Stancu:2020paw}.
Recently a new hidden-charm pentaquark has been found~\cite{Aaij:2020gdg}
--- the $P_{cs}(4459)^0$ --- which we will simply denote as $P_{cs}$
in this work.
This pentaquark has been observed in the $J/\psi \Lambda$ channel, from
which it can be deduced that its quark content is $c\bar{c} s qq$
with $q = u, d$.
Its mass and width are
\begin{eqnarray}
  M_{P_{cs}} = 4458.8 \pm 2.9 {}^{+4.7}_{-1.1} \,{\rm MeV} \, , \,
  \Gamma_{P_{cs}} = 17.3 \pm 6.5 {}^{+8.0}_{-5.7} \,{\rm MeV}\, , \nonumber \\
\end{eqnarray}
but the statistical significance of the signal is merely $3.1\,\sigma$.
Besides, its spin and parity have not been determined yet.
It is also worth noticing that predictions of $P_c$ and $P_{cs}$
pentaquarks~\cite{Wu:2010jy,Wu:2010vk,Yang:2011wz} have been there
long before their eventual observation.

The $P_{cs}$ pentaquark lies a few MeV below the $\bar{D}^* \Xi_c$ threshold
--- $4478.0\,{\rm MeV}$ in the isospin symmetric limit --- suggesting
a strong molecular component~\cite{Chen:2020uif,Liu:2020hcv,Dong:2021juy}.
However there are at least other two nearby thresholds:
the $\bar{D} \Xi_c'$ and $\bar{D} \Xi_c^{*}$ ones at
$4446.0$ and $4513.2\,{\rm MeV}$, respectively (i.e. $32.0$
and $35.2\,{\rm MeV}$ away from the $\bar{D}^* \Xi_c$
threshold).
If the spin of the $P_{cs}$ pentaquark is $J=\tfrac{1}{2}$
($\tfrac{3}{2}$), it will mix with the $\bar{D} \Xi_c'$
($\bar{D} \Xi_c^{*}$) channel, which will result into a molecular picture
more complex than that of the $P_c$ pentaquarks
(i.e. that of a single channel $\bar{D} \Sigma_c$ or
$\bar{D}^*\Sigma_c$ molecule).
Here we will consider how the aforementioned coupled channel dynamics
affects the spectrum of a molecular $P_{cs}$.
If we consider the possible isoscalar $\bar{D} \Xi_c'$ and $\bar{D} \Xi_c^*$
molecular states, one quickly realizes that owing to SU(3)-flavor and
heavy quark spin symmetry (HQSS) it is possible to make
predictions~\cite{Peng:2019wys}.
For the $\bar{D}^* \Xi$ system this is not the case though and
we will have to resort to phenomenology to relate its interaction
with the already known non-strange molecular pentaquarks.
If this is done, the molecular description of the $P_c(4312/4440/4457)$ and
$P_{cs}(4459)$ pentaquarks turns out to be coherent, as we will explain
in the following lines.

The manuscript is organized as follows: in Sect.~\ref{sec:EFT} we briefly
explain the non-relativistic effective field theory we will use to
describe the molecular pentaquarks. In Sect.~\ref{sec:light-heavy}
we discuss the symmetry constraints of the pentaquarks.
Sect.~\ref{sec:pc} is devoted to the power counting of
the coupled channels affecting the $P_{cs}$.
In Sect.~\ref{sec:saturation} we explain how to estimate
the low energy constants of the effective field theory
from meson-exchange saturation. In Sect.~\ref{sec:accidental}
we will show an accidental symmetry between the potentials of 
the $P_c(4440/4457)$ and $P_{cs}$ pentaquarks.
In Sect.~\ref{sec:hyperfine} we discuss the size of the hyperfine splitting
between the $J=\tfrac{1}{2}$, $\tfrac{3}{2}$ $P_{cs}$ pentaquarks.
In Sect.~\ref{sec:decays} we consider the decay of the $P_{cs}$ pentaquark
into $J/\psi \Lambda$ depending on its spin.
Finally, we summarize our conclusions in Sect.~\ref{sec:conclusions}
and explain a few technicalities in Appendices \ref{sec:running}
and \ref{sec:saturation-NN}.



\section{Effective field theory description}
\label{sec:EFT}

Before explaining how symmetries inform the pentaquark spectrum,
first we will briefly explain the effective field theory (EFT)
formalism we follow.
We will describe interactions among heavy hadrons with a non-relativistic
contact-range potential of the type
\begin{eqnarray}
  \langle p' | V | p \rangle = C \, ,
\end{eqnarray}
with $C$ an unknown coupling constant, where this coupling can be further
decomposed into a sum of irreducible components $C = \sum_R \lambda_R C^R$,
with $R$ denoting some quantum-number / representation, $\lambda_R$
some coefficient / operator and $C^R$ the particular coupling
that applies in each case.
This type of contact-range potential often appears in lowest- (or leading-)
order EFT descriptions of hadron-hadron interactions (concrete examples
with full derivations can be found in
Refs.~\cite{AlFiky:2005jd,Mehen:2011yh,Nieves:2012tt,HidalgoDuque:2012pq}
for antimeson-meson molecules and in Ref.~\cite{Liu:2018zzu} 
for pentaquarks).
Of course this is true provided that the one-pion-exchange potential, which is
the longest range piece of the hadron-hadron interaction, is weak and
thus subleading~\cite{Valderrama:2012jv,Lu:2017dvm} (otherwise
it should be included at lowest-order).
The previous contact-range potential is singular though and
has to be regularized, which we do by introducing a regulator
function $f(x)$ and a cutoff $\Lambda$, i.e.
\begin{eqnarray}
  \langle p' | V | p \rangle = C(\Lambda)\,
  f(\frac{p'}{\Lambda})\,f(\frac{p}{\Lambda}) \, , 
\end{eqnarray}
where the coupling now depends on the cutoff $C = C(\Lambda)$.
For the regulator we will choose a Gaussian, $f(x) = e^{-x^2}$, and for
the cutoff we will use the range $\Lambda = 0.5-1.0\,{\rm GeV}$.
Finally this potential is included in a dynamical equation,
such as Schr\"odinger or Lippmann-Schwinger,
for obtaining predictions.
If we choose Lippmann-Schwinger and are interested in poles of
the scattering amplitude, i.e. bound/virtual states or
resonances, we can simply solve
\begin{eqnarray}
  \phi(k) + \int \frac{d^3 p}{(2\pi)^3}\,\langle k | V | p \rangle
  \,\frac{\phi(p)}{M_{\rm th} + p^2/(2 \mu) - M_{\rm mol}} = 0,\notag\\ \, 
  \label{eq:BSE}
\end{eqnarray}
where $\phi$ is the vertex function, which is defined as the 
the wave function $\Psi$ times the propagator
($\phi(p) = [M_{\rm th} + p^2/(2\mu) - M_{\rm mol}]\,\Psi(p)$),
$V$ the potential, $M_{\rm th}$ the mass of the threshold
(i.e. the sum of the masses of the two hadrons comprising
a molecular candidate), $\mu$ their reduced mass and
$M_{\rm mol}$ the mass of the hadronic molecule we want to predict.



\section{Light-flavor and heavy-quark symmetries}
\label{sec:light-heavy}

Symmetry constrains the potential binding the molecular pentaquarks.
If we begin by considering the three known $P_c$ pentaquarks,
in the molecular picture they are thought to be $\bar{D} \Sigma_c$
and $\bar{D}^* \Sigma_c$ bound states.
From the SU(3)-flavor perspective the $P_c$'s are composed of
a triplet charmed antimeson and a sextet charmed baryon, which together
can couple into the octet and decuplet representations of SU(3),
i.e. $3 \otimes 6 = 8 \oplus 10$.
The flavor structure of the potential is thus
\begin{eqnarray}
  V(\bar{H}_c S_c) = \lambda^O C^O + \lambda^D C^D \, ,
\end{eqnarray}
with $H_c= D, D^*$ or $D_s, D_s^*$ and $S_c = \Sigma_c, \Sigma_c^*$,
$\Xi_c', \Xi_c^*$ or $\Omega_c, \Omega_c^*$ representing an arbitrary
charmed meson or baryon, $C^O$ and $C^D$ the octet and decuplet
couplings and $\lambda^O$ and $\lambda^D$ a coefficient
that depends on the particular antimeson-baryon
configuration considered (they are explained
in detail in Ref~\cite{Peng:2019wys}).

From the HQSS perspective the potential between two heavy hadrons can only
depend on the spin of the light quarks inside them.
For the triplet charmed meson and sextet charmed baryon the light-spins are
$S_L = \tfrac{1}{2}$ and $S_L = 1$, respectively, which couple to
$\tfrac{1}{2} \otimes 1 = \tfrac{1}{2} \oplus \tfrac{3}{2}$.
However it is more compact to express the light-quark spin structure of
the potential in terms of light-spin operators:
\begin{eqnarray}
  V(\bar{H}_c S_c) = C_a + C_b\,\vec{\sigma}_L \cdot \vec{S}_L \, ,
  \label{eq:HS}
\end{eqnarray}
with $C_a$ and $C_b$ couplings that represent the spin-independent and
spin-dependent pieces of the potential, respectively, and
$\vec{\sigma}_L$ and $\vec{S}_L$ the spin-operators
for the light-spin degrees of freedom within
the charmed meson and baryon (for the notation in terms of
light-spin check for instance Ref.~\cite{Valderrama:2019sid},
while the channel-by-channel potential can be found in Ref.~\cite{Liu:2018zzu}).

From the SU(3)-flavor and HQSS structure we have just explained it is already
possible to derive the  existence of $\bar{D} \Xi_c'$ and $\bar{D} \Xi_c^*$
molecular states~\cite{Peng:2019wys}.
First we notice that the standard molecular interpretation of
the $P_c(4312)$ pentaquark is that it is a $I=\tfrac{1}{2}$
$\bar{D} \Sigma_c$ bound state.
Thus the decomposition of the $P_c(4312)$ potential is
\begin{eqnarray}
  V(\bar{D} \Sigma_c, I=\frac{1}{2}) = C^O_a \, , 
\end{eqnarray}
i.e. the octet SU(3)-representation and the HQSS part of
the potential that is independent of the spin of the light-quarks.
Any other molecular pentaquark with the same decomposition will
have the same potential as the $P_c(4312)$ and consequently,
will be likely to have a similar binding energy.
Among these pentaquarks we have the $I=0$ $\bar{D} \Xi_c'$ and $\bar{D} \Xi_c^*$
systems, for which the potential reads
\begin{eqnarray}
  V(\bar{D} \Xi_c', I=0) = V(\bar{D} \Xi_c^*, I=0) = C^O_a \, , 
\end{eqnarray}
where again only the octet, spin-independent piece of
the contact-range potential ($C^O_a$) is involved.
From now on we will simply write $C_a = C_a^O$, as the octet configuration is
the only one we are considering in this work.

If we determine this coupling from the $P_c(4312)$, we can predict
the masses of the $\bar{D} \Xi_c'$ and $\bar{D} \Xi_c^*$ molecules,
i.e. the $P_{cs}'$ and $P_{cs}^*$ pentaquarks, with the formalism
we already described.
The result happens to be
\begin{eqnarray}
M(P_{cs}') &=& 4436.7 \,(4436.1)\,{\rm MeV} \, , \\
M(P_{cs}^*) &=& 4503.6 \,(4502.7)\,{\rm MeV} \, ,
\end{eqnarray}
for $\Lambda = 0.5 (1.0)\,{\rm GeV}$, where similar predictions can be found
in Refs.~\cite{Peng:2019wys,Xiao:2019gjd}.

If we now consider the $P_{cs}$, its most natural molecular interpretation
is $\bar{D}^* \Xi_c$.
This two-body system is not connected to $\bar{D} \Xi_c'$, $\bar{D} \Xi_c^*$
and $\bar{D} \Sigma_c$ neither by SU(3)-flavor nor HQSS symmetries.
From SU(3)-flavor symmetry, the $P_{cs}$ pentaquark contains a triplet
charmed antimeson and antitriplet charmed baryon and is a combination
of a singlet and an octet, i.e. $3 \otimes \bar{3} = 1 \oplus 8$.
The concrete flavor structure of the potential is unessential though,
as we are only considering the $I=0$, $S=-1$ sector (i.e. $\bar{D}^* \Xi_c$).
Regarding HQSS, the antitriplet charmed baryon contains a diquark
with $S_L = 0$, from which we expect a trivial light-spin structure
owing to $\tfrac{1}{2} \otimes 0 = \tfrac{1}{2}$.
The potential reads
\begin{eqnarray}
  V(\bar{H}_c T_c) = D_a  \, , \label{eq:HT}
\end{eqnarray}
with no spin dependence whatsoever and $T_c = \Lambda_c, \Xi_c$ representing
a generic antitriplet charmed baryon.
In addition to this, the $\bar{H}_c T_c$ and $\bar{H}_c S_c$ systems
can couple by means of a transition potential of the type
\begin{eqnarray}
  V(\bar{H}_c T_c - \bar{H_c} S_c) = E_b \, \vec{\sigma}_L \cdot \vec{\epsilon}_L
  \, , \label{eq:HT-HS}
\end{eqnarray}
with $E_b$ a coupling, $\vec{\sigma}_L$ the spin-operator for the light-quark
within the charmed meson and $\vec{\epsilon}_L$ the polarization vector
of the light-diquark in the sextet charmed baryon.
The couplings $D_a$ and $E_b$ can be further decomposed in isospin and flavor
representations, but this is not necessary for the set of molecules
we are considering.
Putting the pieces together for the $\bar{D}^* \Xi_c$, if
we consider the coupled channel bases
$\mathcal{B}(J={\frac{1}{2}}) = \{ \bar{D} \Xi_c' , \bar{D}^* \Xi_c \}$
and $\mathcal{B}(J={\frac{3}{2}}) = \{ \bar{D}^* \Xi_c , \bar{D} \Xi_c^* \}$
we will have the following potentials:
\begin{eqnarray}
  V(P_{cs}, J=\frac{1}{2}) &=&
  \begin{pmatrix}
    C_a & {E_b} \\
    {E_b} & D_a
  \end{pmatrix} \, , \\
  V(P_{cs}, J=\frac{3}{2}) &=& 
  \begin{pmatrix}
    D_a & E_b \\
    E_b & C_a
  \end{pmatrix} \, .
\end{eqnarray}
By including these potentials in a bound state equation such as
the coupled-channel extension of Eq.~(\ref{eq:BSE})
we can calculate the mass of the $P_{cs}$.



\section{Power counting and coupled channel dynamics}
\label{sec:pc}

EFTs are expected to be power series in terms of the expansion parameter
$(Q/M)$, where $Q$ and $M$ represent the characteristic low and high
energy scales of the system, respectively.
For molecular pentaquarks $Q$ is of the order of the pion mass
($m_{\pi} \simeq 140\,{\rm MeV}$) or the wave number of
the bound state (i.e. $\gamma = \sqrt{2 \mu B_2} \sim 206\,{\rm MeV}$
for the $P_{cs}$ as a $\bar{D}^* \Xi_c$ molecule), while
$M$ will be of the order of the vector meson mass
($m_{\rho} \simeq 770\,{\rm MeV}$).
This suggests the expansion parameter
\begin{eqnarray}
  \frac{Q}{M} \sim \frac{\sqrt{2 \mu B_2}}{m_{\rho}} \sim 0.27 \, ,
\end{eqnarray}
where we have identified the wave number $\gamma$ with the light scale $Q$.
We will now compare this number with the expected size of
coupled channel effects.

If we are interested in the mass difference between the $J=\tfrac{1}{2}$
and $\tfrac{3}{2}$ $\bar{D}^* \Xi_c$ bound states,
i.e. the  hyperfine splitting,
the relevant coupled channels are of the $\bar{H}_c T_c$-$\bar{H}_c S_c$ type,
i.e. Eq. (\ref{eq:HT-HS}), which can break the spin degeneracy.
There are $\bar{H}_c T_c$-$\bar{H}_c T_c$ coupled channel effects too
(e.g. $\bar{D}_s^* \Lambda_c$-$\bar{D}^* \Xi_c$),
but they do not generate a dependence on the light spin.
For the coupled channel dynamics relevant to the $\bar{D}^* \Xi_c$ system
(independently of whether they generate spin dependence),
their expected size with respect to the diagonal interaction is~\cite{Valderrama:2012jv,Lu:2017dvm}
\begin{eqnarray}
  &&{\left( \frac{Q}{\Lambda_{CC}} \right)}^2 \sim \frac{B_2}{\Delta_{CC}}
  \sim 0.60 \, , \, 0.54 \, , \, 0.24 \, , \,  0.14 \, , \, 0.11
  \, ,
  \nonumber \\
  && \qquad \qquad \quad \mbox{for} \,\,
  \bar{D} \Xi_c' \, , \, \bar{D} \Xi_c^* \, , \, \bar{D}^*_s \Lambda_c
  \, , \,  \bar{D} \Xi_c \, , \, \bar{D}^* \Xi_c^*
  \, , \nonumber \\
\end{eqnarray}
respectively, where $B_2 = 19.2\,{\rm MeV}$ is the binding energy of
a molecular $P_{cs}$ and $\Delta_{CC}$ the mass gap of the listed
coupled channels.
This indicates that only the $\bar{D} \Xi_c'$ and $\bar{D} \Xi_c^*$ channels
are expected to be larger than the size of subleading corrections.
The next channel in importance, $\bar{D}^*_s \Lambda_c$, does not break
spin degeneracy, as previously mentioned, and in addition
its size is subleading.
Finally, though the $\bar{D}^* \Xi^*$ channel will indeed contribute to the
hyperfine splitting, its size is strongly suppressed with respect to
$\bar{D} \Xi_c'$ and $\bar{D} \Xi_c^*$ and thus we will not take it
into account.

\begin{figure}
  \begin{center}
    \epsfig{figure=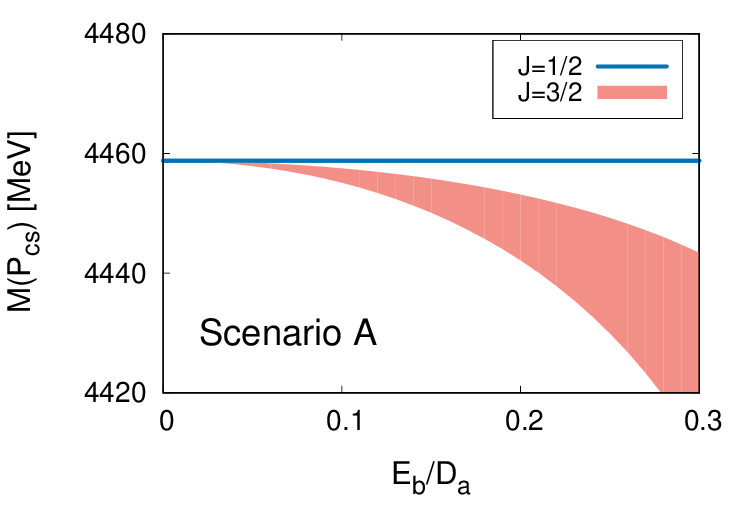,
      height=5.0cm, width=8.0cm}
    \epsfig{figure=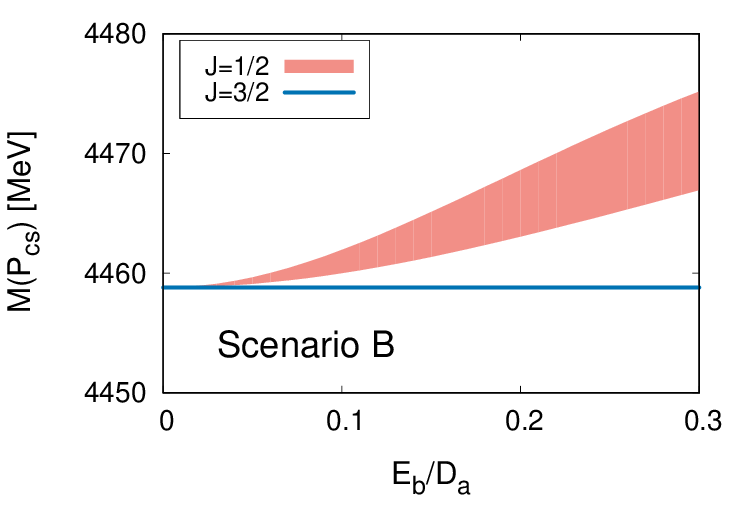,
      height=5.0cm, width=8.0cm}
    \epsfig{figure=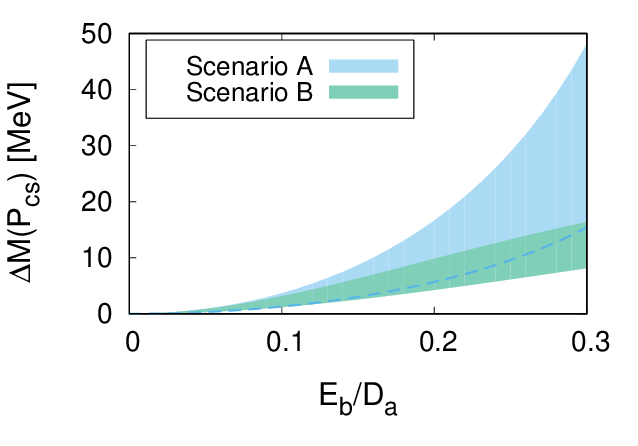,
      height=5.0cm, width=8.0cm}
\end{center}
\caption{
  Masses and hyperfine splitting of the $J=\tfrac{1}{2}$ and $\tfrac{3}{2}$
  molecular $P_{cs}$ pentaquarks.
  For $J=\tfrac{1}{2}$ ($\tfrac{3}{2}$) we include the coupled channel dynamics
  $\bar{D} \Xi_c'$-$\bar{D}^* \Xi_c$ ($\bar{D}^* \Xi_c$-$\bar{D} \Xi_c^*$),
  which are expected to be ${\rm LO}$ effects in the EFT description.
  This description contains three independent couplings $C_a$, $D_a$, $E_b$,
  where the first one ($C_a$) is determined from the $P_c(4312)$ pentaquark
  and SU(3)-flavor symmetry.
  For the other two couplings we do as follows: scenario $A$ ($B$) assumes
  that the observed $P_{cs}$ pentaquark is the $J=\tfrac{1}{2}$
  ($\tfrac{3}{2}$) configuration; then for a given $E_b / D_a$ ratio
  we determine $D_a$ from the $P_{cs}$ mass.
  Finally, we show the hyperfine splitting
  $\Delta M(P_{cs}) = M(P_{cs}, J=\tfrac{1}{2}) - M(P_{cs}, J=\tfrac{3}{2})$
  in both scenarios.
}
\label{fig:hyperfine-splitting}
\end{figure}

For analyzing the possible impact of the coupled channel dynamics,
we will do the following calculation
\begin{itemize}
\item[(i)] Consider the $P_{cs}$ pentaquark to be a $J=\tfrac{1}{2}$ or
  $\tfrac{3}{2}$ molecule, which in analogy with Ref.~\cite{Liu:2019tjn}
  we will call scenarios A and B, respectively.
\item[(ii)] Consider different $E_b / D_a$ coupling ratios: with this ratio
  fixed, the $D_a$ coupling can be determined from the $P_{cs}$ pentaquark
  (and the $C_a$ from the $P_c(4312)$ one). Then we check
  how the hyperfine splitting changes with this ratio.
\end{itemize}
The result of these calculations is shown in Fig.~\ref{fig:hyperfine-splitting}
for scenarios A and B and a cutoff $\Lambda = 0.5-1.0\,{\rm GeV}$.
The hyperfine splitting grows quickly with the $E_b / D_a$ ratio and
it is sizable even for small ratios.
This can be understood from the power counting of contact-range
theories~\cite{vanKolck:1998bw},
in which a coupling generating a bound state near threshold is fine-tuned,
thus explaining how the effect of a comparatively small $E_b$ is amplified
by the fact that $D_a$ can generate a molecular $P_{cs}$.
We also notice that the for the same $E_b/D_a$ ratio the hyperfine splitting
will be considerably larger in scenario $A$, which has to do with the fact
that $D_a$ is also larger in this scenario: coupled-channel dynamics
require $M(J=\tfrac{1}{2}) > M(J=\tfrac{3}{2})$, which in turn forces
$D_a$ to be larger in scenario A if $E_b \neq 0$.

However, without being able to estimate the $E_b / D_a$ coupling
it will be not possible to know the hyperfine splitting.
From power counting arguments the size of each of these couplings
will be~\cite{vanKolck:1998bw}
\begin{eqnarray}
  | D_a^{(R)} | \propto \frac{2 \pi}{\mu \sqrt{2 \mu B_2}} \quad \mbox{and} \quad
  | E_b^{(R)} | \propto \frac{2 \pi}{\mu M} \, , \label{eq:coupling-ratio}
\end{eqnarray}
where the superscript ${}^{(R)}$ refers to a {\it renormalized} coupling
(we elaborate below): $D_a^{(R)}$ is said to be {\it enhanced}
(i.e. its size is larger than expected owing to the existence of
a bound state close to threshold), while $E_b^{(R)}$ is {\it natural}
(i.e. its size can be determined from standard or naive
dimensional analysis arguments).
The renormalized couplings $D_a^{(R)}$ and $E_b^{(R)}$ (which is the type of
couplings for which the arguments of Ref.~\cite{vanKolck:1998bw}
were originally developed) refer {\it loosely speaking}
to the parts of the couplings that do not
depend on the cutoff.
But here we are working instead with the {\it bare (or running)}
couplings $D_a = D_a(\Lambda)$ and $E_b = E_b(\Lambda)$,
which explicitly depend on the cutoff.
Nonetheless the previous power counting estimates apply
to the bare couplings for specific cutoff ranges:
  (i) for couplings of natural size (e.g. $E_b$), this will be
  the case irrespectively of whether the cutoff is soft
  ($\Lambda \sim Q$) or hard ($\Lambda \sim M$), while
  (ii) for couplings of unnatural size (e.g. $D_a$) the enhancement
  ideally requires a soft cutoff $\Lambda \sim Q$, with the size of
  the coupling reverting to its natural size as the cutoff
  becomes harder~\cite{Valderrama:2016koj,Epelbaum:2017byx}.
  However, in the molecular pentaquarks the separation of scales is
  far from perfect: as explained in Appendix \ref{sec:running}
  for the $P_{cs}$ pentaquark and the Gaussian regulator
  we use here, $D_a(\Lambda)$ coincides with its power counting estimation
  for $\Lambda \sim 0.9\,{\rm GeV}$ (i.e. within the cutoff range we use).
  We thus expect that in a first approximation the previous relations
  will hold in the $\Lambda = (0.5-1.0)\,{\rm GeV}$ range, i.e.
  \begin{eqnarray}
    \frac{E_b(\Lambda)}{D_a(\Lambda)} \simeq \frac{E_b^{(R)}}{D_a^{(R)}} \, ,
  \end{eqnarray}
  from which we get
$E_b / D_a \sim Q / M \sim 0.27$, yielding an estimated hyperfine
splitting of $\Delta M^A \sim 12-35$ and $\Delta M^B \sim 7-15\,{\rm MeV}$
in scenarios $A$ and $B$, respectively, where there is still a
  noticeable cutoff dependence.
It is nonetheless possible to improve over the previous picture by
including a renormalization factor to better connect the bare and
running couplings:
  \begin{eqnarray}
    \frac{E_b(\Lambda)}{D_a(\Lambda)} \simeq \mathcal{F}(\frac{Q}{\Lambda})\,
    \frac{E_b^{(R)}}{D_a^{(R)}} \, , \label{eq:PC-running}
  \end{eqnarray}
  where we discuss the derivation of the factor $\mathcal{F}$
  in Appendix \ref{sec:running}.
  Concrete calculations show that $\mathcal{F} = (1.47-0.91)$
  in the $\Lambda = 0.5-1.0\,{\rm GeV}$ cutoff window used
  in this work, leading to the hyperfine splittings
  $\Delta M^A \sim 27-40$ and $\Delta M^B \sim 12-13\,{\rm MeV}$
  in scenarios $A$ and $B$, respectively, which happen to display
  less cutoff dependence (though they are still of the same order
  as our original estimation).
  It is worth noticing that the reason behind these elaborations
  is that we are applying power counting arguments to non-observable
  quantities (which are allowed to have a strong dependence
  on the cutoff).
  
  Actually, besides the standard $P_{cs}(4459)$ single peak found
  in the $J/\psi \Lambda$ invariant mass distribution,
  the LHCb collaboration also reports a second possible
  two-peak solution~\cite{Aaij:2020gdg}
  involving two $P_{cs}$ pentaquarks with masses 
  \begin{eqnarray}
    M(P_{cs1}) &=& 4454.9 \pm 2.7 \, {\rm MeV} \, , \label{eq:Pcs1} \\
    M(P_{cs2}) &=& 4467.8 \pm 3.7 \, {\rm MeV} \, , \label{eq:Pcs2}
  \end{eqnarray}
  which, if they are both to be interpreted primarily as $\bar{D}^* \Xi_c$ bound
  states, will result in the hyperfine splitting
  $\Delta M = 12.9 \pm 4.6 \,{\rm MeV}$.
  In principle this is compatible with scenario $B$.
  But both scenarios $A$ and $B$ use the standard single peak solution
  as the reference input.
  Had we determined the $D_a$ and $E_b$ couplings from the two-peak
  solution instead, then the $E_b/D_a$ ratio would have been
  \begin{eqnarray}
    \frac{E_b}{D_a} = (0.37-0.22) \, , 
  \end{eqnarray}
  for the cutoff range $\Lambda = 0.5-1.0\,{\rm GeV}$.
  This ratio is in fact compatible with the power counting estimation of
  $Q/M \sim 0.27$.
In the following lines we will resort to phenomenological information
for further elucidating the $E_b/D_a$ ratio.

\section{Meson exchange saturation}
\label{sec:saturation}

The problem we have is that there are three couplings ($C_a$, $D_a$, $E_b$)
of which we can only determine two ($C_a$ from the $P_c(4312)$ and
$D_a$ or $E_b$ from the $P_{cs}(4459)$).
Yet, if we use phenomenology it might be possible to find relations among
these couplings and thus determine the three of them.
In particular we will focus on light-meson saturation, i.e.
the idea that the contact-range couplings of a given EFT are saturated
by the exchange of light mesons~\cite{Ecker:1988te,Epelbaum:2001fm}.
Here we choose the novel saturation procedure of Ref.~\cite{Peng:2020xrf},
which we explain below.

Standard saturation maps the finite-range S-wave potential generated by
the exchange of a light-meson, $V_M({\vec{q}\,}^2)$
(with $\vec{q} = \vec{p}\,' - \vec{p}$ the exchanged momentum, where
for S-wave we can express the potential as a function of ${\vec{q}\,}^2$),
into a contact-range coupling $C^{\rm sat}$ by taking the limit
\begin{eqnarray}
  C^{\rm sat\,(standard)}(\Lambda \sim m)
  \propto \lim_{\vec{q}^2 \to 0} V_M(\vec{q}^2) \, ,
  \label{eq:standard-sat}
\end{eqnarray}
which is expected to work for $\Lambda$ close to
the mass $m$ of the exchanged meson.
However, if the potential vanishes in this limit,
we will obtain $C^{\rm sat} = 0$.
For instance, the potentials
\begin{eqnarray}
  V_M = - g_Y^2\,\frac{1}{m^2 + \vec{q}^2}
  \quad \mbox{and} \quad
  V_M' = + g_Y^2\,\frac{\vec{q}^2}{m^2}\frac{1}{m^2 + \vec{q}^2} \, ,
\end{eqnarray}
generate exactly the same finite-range potential in r-space,
namely the Yukawa potential
\begin{eqnarray}
  V_M^{(')}(r) = - g_Y^2 \frac{e^{-m r}}{4 \pi r} \quad \mbox{for $r \neq 0$} \, ,
\end{eqnarray}
where the difference between the two is a distribution
\begin{eqnarray}
  V_M'(r) - V_M(r) = g_Y^2\,m\,\delta^{(3)}(m\,\vec{r}) \, .
\end{eqnarray}
Of course, at this point we have to discuss the impact of form-factors,
  which modify the light-meson exchange potentials as follows
  \begin{eqnarray}
    V_M^{(')}(\vec{q}; \Lambda_M) = V_M^{(')}(\vec{q}) F_M^2(\vec{q}, \Lambda_M)
    \, ,
  \end{eqnarray}
  where $F_M$ is the aforementioned form-factor (with the most used
  parametrizations being of the multipolar type) while $\Lambda_M$ is
  the form-factor cutoff, which should not be confused
  with the EFT cutoff $\Lambda$.
  For a local form-factor the resulting r-space potentials will be also local,
  and the Dirac-delta will acquire a finite-size:
  \begin{eqnarray}
    \delta^{(3)}(m\,\vec{r}) \to \delta_{F}^{(3)}(m\,\vec{r}; \frac{m}{\Lambda_M})
    \, , 
  \end{eqnarray}
  where $\delta_F^{(3)}$ represents a Dirac-delta that has been already
  smeared out by the form-factor.
  Yet, the characteristic scale $\Lambda_{M}$ of these finite-range effects is
expected to be larger than the mass of the exchanged meson,
i.e. $\Lambda_{M} > m$ (otherwise the effect of said exchanged
light-meson will be washed out by the form factors).
For instance, for the Bonn-B potential~\cite{Machleidt:1987hj}
$\Lambda_{\sigma} = 1.9-2.0\,{\rm GeV}$ and
$\Lambda_{\rho/\omega} = 1.85\,{\rm GeV}$
for the $M = \sigma$ and $\rho$/$\omega$ mesons, respectively, while
for the CD-Bonn potential~\cite{Machleidt:2000ge} we have
$\Lambda_{\sigma} = 2.5\,{\rm GeV}$, $\Lambda_{\rho} = 1.31\,{\rm GeV}$
and $\Lambda_{\omega} = 1.5\,{\rm GeV}$.

Thus for the range of cutoffs in which saturation is expected to work
we have $(\Lambda \sim m) < \Lambda_{F}$, which implies that
the previous Dirac-delta is unimportant:
independently of whether the potential is derived from derivative
interactions ($V_M'$) or not ($V_M$), the potentials at $m r \sim 1$
will be similar and thus the saturation of the couplings
should follow suit.
That is, if the renormalization scale is similar to the exchanged meson mass,
the potentials $V_M$ and $V_M'$ are expected to lead to approximately
the same saturated coupling.
This is achieved with the convention
\begin{eqnarray}
  C^{\rm sat\,(new)}(\Lambda \sim m) \propto \frac{1}{m^2}\,
  \underset{{\vec{q}^2 \to -m^2}}{\rm Res} V_M^{(')}(\vec{q}^2) \, ,
  \label{eq:modifed-sat}
\end{eqnarray}
i.e. by extracting the residue of the potential at $\vec{q}^2 = -m^2$,
which effectively recovers the expectations from the $m r \sim 1$
behavior of the potential.
Ref.~\cite{Peng:2020xrf} explicitly checked this method with
the one-pion-exchange potential (with an arbitrary coupling
strength) as a specific example.
In Appendix~\ref{sec:saturation-NN} we include a detailed comparison
  between the standard saturation procedure of Ref.~\cite{Epelbaum:2001fm}
  and the one presented here for the particular case of
  the nucleon-nucleon system, which indicates that both
  saturation methods yield comparable results.

Now, if we consider the scalar meson $\sigma$, in the non-relativistic limit
it generates a spin-independent potential which can contribute
to the saturation of $C_a$ and $D_a$ (but not $C_b$ or $E_b$):
\begin{eqnarray}
  V_S(\vec{q}) &=& -\frac{g_{\sigma i} g_{\sigma j}}{m_{S}^2 + {\vec{q}\,}^2} \, ,
\end{eqnarray}
where $m_S$ the mass of the sigma meson and $g_{\sigma i}$ its coupling, and
the indices $i, j = 1,2,3$ referring to the $\bar{D}^{(*)}$, $\Xi_c^{('/*)}$
and $\Xi_c$, respectively.
Independently of the saturation method we will obtain
\begin{eqnarray}
  F_a^{\rm sat(S)}(\Lambda \sim m_S) &\propto& -\frac{g_{\sigma i} g_{\sigma j}}{m_S^2} \, , 
\end{eqnarray}
with $F_a = C_a, D_a$ the generic name for the spin-independent couplings.
The proportionality constant is in principle unknown,
but we will assume it to be similar for all the couplings.
We remind that saturation is expected to work for cutoffs close to the mass
of the meson being exchanged, $\Lambda \sim m_S$ in this case.

The vector mesons ($\rho$ and $\omega$) generate a more complex potential
which can be expanded in a multipole expansion similar to the one we have
for electromagnetic interactions.
There are electric- and magnetic-type components (indicated by
the subscripts $a$ and $b$)
\begin{eqnarray}
  V_V &=& V_{Va} + V_{Vb} \, ,
\end{eqnarray}
with
\begin{eqnarray}
  V_{Va} &=& +(1 + \vec{\tau}_1 \cdot \vec{\tau}_2)\,
  \frac{g_{V i} g_{V j}}{m_V^2 + {\vec{q}\,}^2} \, , \\
  V_{Vb} &=& -(1 + \vec{\tau}_1 \cdot \vec{\tau}_2)\,
  \frac{f_{V i} f_{V j}}{6 M^2}\,\mathcal{O}_{L12}\,
  \frac{{\vec{q}\,}^2}{m_V^2 + {\vec{q}\,}^2} + \dots\, ,
\end{eqnarray}
and $O_{L12} = 0$, $\vec{\sigma}_L \cdot \vec{\epsilon}_L$ or
$\vec{\sigma}_{L} \cdot \vec{S}_{L}$ for $\bar{H}_c T$,
$\bar{H}_c T_c$-$\bar{H}_c S_c$ and $\bar{H}_c S_c$,
respectively, where the dots represent S-to-D-wave components
(we assume they will not appreciably contribute to
the saturation of the EFT couplings).
In the vector-exchange potential the $i, j$ indices refer to the different
hadrons involved, $g_{Vi}$ to the electric-type couplings,
$f_{Vi}$ to the magnetic-like ones, $m_V$ to the vector meson mass
and $M$ a typical hadronic mass scale.
The $V_{Va}$ component contribute to the $C_a$ and $D_a$ couplings as
\begin{eqnarray}
  F_a^{\rm sat(V)}(\Lambda \sim m_V)
  &\propto& (1 + \vec{\tau}_1 \cdot \vec{\tau}_2)\,
  \frac{g_{V i} g_{V j}}{m_V^2} \, , 
\end{eqnarray}
where $F_a = C_a$, $D_a$, with no difference between the standard and
modified saturation procedures, i.e. Eqs.~(\ref{eq:standard-sat})
and ~(\ref{eq:modifed-sat}).
For the $V_{Vb}$ component to contribute to the $C_b$ and $E_b$ couplings
in a non-trivial way we have to use the modified saturation procedure
of Ref.~\cite{Peng:2020xrf} (i.e. Eq.~(\ref{eq:modifed-sat}))
in which case we obtain
\begin{eqnarray}
  F_b^{\rm sat(V)}(\Lambda \sim m_V)
  &\propto& (1 + \vec{\tau}_1 \cdot \vec{\tau}_2)\,
  \frac{f_{V i} f_{V j}}{6 M^2} \, ,
\end{eqnarray}
where $F_b = C_b, E_b$.
Here we notice that with the standard saturation method we would have arrived
to $C_b = E_b = 0$.
This is what happens for instance in Refs.~\cite{Xiao:2019aya,Xiao:2019gjd},
which in principle consider the complete set of ${\rm LO}$ interactions
we have here, i.e. Eqs.~(\ref{eq:HS}), (\ref{eq:HT}) and (\ref{eq:HT-HS}),
but set the $F_b$-type couplings to zero leading to pentaquark predictions
without sizable hyperfine splittings.
Yet, as we will see, from the point of view of phenomenology the choice of
Refs.~\cite{Xiao:2019aya,Xiao:2019gjd} is completely justified:
the size of the $F_b$-type of coupling is expected to be considerably smaller
than the $F_a$-type, which prompts the approximation $F_b = 0$.
In contrast from the EFT point of view the inclusion of the $F_b$ couplings is
justified in terms of power counting.

The only thing left is to determine the couplings: for the $\sigma$ we will
use the linear-sigma model~\cite{GellMann:1960np} and the quark model
(as used in Ref.~\cite{Riska:2000gd}),
from which we get that the coupling of the sigma to the nucleon
is $g_{\sigma NN} = \sqrt{2} M_N / f_{\pi} \sim 10.2$
(for the $f_{\pi} \simeq 132\,{\rm MeV}$ normalization).
For the charmed mesons with only one light-quark we end up with
$g_{\sigma 1} = g_{\sigma NN} / 3 \simeq 3.4$, while for the charmed baryons
with two light-quarks we have
$g_{\sigma 2} = g_{\sigma 3} = 2\,g_{\sigma NN} / 3 \simeq 6.8$.
We note that here we have assumed the same coupling of the sigma
with the $q = u,d,s$ quarks, as happens for instance
in the quark model of Ref.~\cite{Vijande:2009pu},
where we notice that this pattern can either arise from
  a singlet sigma (not necessarily a realistic assumption) or
  alternatively from a negligible coupling of the octet
  component of the sigma to the light quarks, see for instance
  Ref.~\cite{Yan:2021tcp} for a more detailed discussion.
In either case,
this runs counter to the standard expectation that the strange and
non-strange components of light mesons would decouple (owing to the OZI rule),
leading in the case of the sigma to a vanishing coupling to the strange quark.
However it has been argued that the OZI rule does not work so well
in the $0^{++}$ sector~\cite{Geiger:1992va,Lipkin:1996ny,Isgur:2000ts,Meissner:2000bc}, i.e. a non-singlet sigma might potentially
have a non-negligible coupling with the strange quark.
Besides, the singlet and octet mixing angle of the sigma might very well be
far away from the angle decoupling strange and non-strange
components~\cite{Oller:2003vf}.
Yet, even though the sigma coupling to the strange quark might not be
necessarily suppressed, it might not be as strong as with the light-quarks.
This can generate SU(3)-flavor breaking effects which we will discuss later,
where the naive expectation will be that the contribution of the sigma
to the contact-range couplings will be weaker in the $\bar{D} \Xi_c^{'}$
and $\bar{D}^* \Xi_c$ molecules than in the $\bar{D} \Sigma_c$ one.
For further details we refer to the discussion around
Eq.~\ref{eq:ratio-std-sigma}, where we notice that the previous expectation
seems to be challenged by the experimental information currently
available~\footnote{In particular
  $D_a$ (as obtained from the $P_{cs}(4459)$) seems to be more attractive
  than $C_a$ (as obtained from the $P_c(4312)$), see Eq.~(\ref{eq:rat-1}),
  which runs counter with the expectation that sigma exchange
  should be weaker in the former case. Yet more accurate experimental
  measurements are needed to confirm whether this is really the case.}.

For the vector mesons, we have electric- and magnetic-like couplings,
for which we will resort to Sakurai's universality and vector meson
dominance~\cite{Sakurai:1960ju,Kawarabayashi:1966kd,Riazuddin:1966sw}, 
i.e. the mixing of the neutral vector mesons with the electromagnetic current,
which can be encapsulated in the substitution rules~\cite{Liu:2019zvb,Peng:2021hkr}
\begin{eqnarray}
  \rho^3_{\mu} \to \frac{1}{\beta}\,\frac{e}{2 g}\,A_{\mu} \quad \mbox{and} \quad
  \omega_{\mu} \to \frac{1}{\beta}\,\frac{e}{6 g}\,A_{\mu} \, ,
\label{eq:vector-em-mixing}
\end{eqnarray}
where $e$ is the proton charge, $g = m_V / 2 f_{\pi} \simeq 2.9$ the universal
rho coupling constant (with $m_V$ the vector meson mass and
$f_{\pi} \simeq 132\,{\rm MeV}$ the pion weak decay constant),
$A_{\mu}$ the photon field, $\rho^3_{\mu}$ the neutral rho field (the superscript
refers to the isospin index, $a = 3$ for the neutral component),
$\omega_{\mu}$ the omega field and $\mu$ a Lorentz index.
The parameter $\beta$ indicates the degree of vector meson dominance:
if the electromagnetic couplings of the heavy hadrons were to be completely
dominated by the vector mesons, we will have $\beta = 1$.
Otherwise it will be $\beta < 1$.
Usually $\beta \simeq 0.9$ is estimated~\cite{Isola:2003fh,Liu:2011xc,Dong:2019ofp}
(which can be traced back
to the ratio of the couplings required for the $\rho \to \gamma \gamma$
and $\rho \to \pi \pi$ processes, i.e. $\sqrt{2} f_{\pi} / f_{\rho} \sim 0.9$).
By applying these substitution rules to the interaction of the heavy hadrons
with the vector mesons, we will get the electromagnetic interaction
for the light quarks within the heavy hadrons and from this,
we can determine the $g_{Vi}$ and $f_{Vi}$ couplings.
In practical terms this means that $g_{Vi}$ and $f_{Vi}$ are proportional to
the light-quark contribution of the electric charge and magnetic moments
of the heavy hadrons, respectively.
For the E0 couplings we get $g_{V 1} = g_{V 2} = g_{V 3} = \beta g$.
For the M1 couplings, we will first make the decomposition $f_{V} = \kappa_V g_V$
and use the choice $M = m_N$ for the mass scale with $m_N \simeq 940\,{\rm MeV}$
the nucleon mass.
From this $\kappa_V = \frac{3}{2} (\mu_u / \mu_{\rm n.m.})$,
with $\mu_u / \mu_{\rm n.m.}$ the magnetic moment of the u-quark
within light-diquark pair inside the $\Xi_c^{+}$/$\Xi_c^{('/*)+}$
charmed baryon expressed in units of the nuclear magneton ($\mu_{\rm n.m.}$).
If we make use of the quark model a second time, we obtain
$\kappa_{V1} = \kappa_{V2} = \frac{3}{2} \mu_u / \mu_{\rm n.m.} \simeq 2.9$
for the charmed antimeson and sextet strange charmed baryon.
For the antitriplet charmed baryon we have instead $\kappa_{V3} = 0$,
which is a consequence of the two light-quarks within the $\Xi_c$ baryon
being in a spin-0 configuration.
The $E_b$ coupling involves a M1 antitriplet to sextet transition
for which $\kappa_V (\bar{3} \to 6) = \frac{3}{2} \, \mu(\bar{3} \to 6) / \mu_{\rm n.m.} = \frac{3}{2} (\mu_u / \mu_{\rm n.m.}) \simeq 2.9$,
where $\mu(\bar{3} \to 6)$ refers to the u-quark magnetic moment in the
$\bar{3} \to 6$ light-diquark transition~\footnote{For translating the magnetic
  moment of the light-diquark into the one for the heavy baryons we can use
  the relations $\mu_{q}(\Xi_c') = \tfrac{2}{3}\,\mu_{q}(6)$ and
  $\mu_{q}(\Xi_c \to \Xi_c') = \pm\mu_{q}(\bar{3} \to 6) / \sqrt{3}$,
  with $\mu_q(X)$ the magnetic moment of a particular light-quark $q$
  within ``$X$'' ($X$ being a baryon or a light-diquark in the antitriplet
  or sextet configuration) and with the sign of the $\bar{3} \to 6$
  transition depending on the ordering of the light quarks in the $\bar{3}$
  flavor wavefunction (which is flavor antisymmetric).}
within $\Xi_c^{+} \to \Xi_c^{(',*)+}$.

With the previous couplings and setting $m_S \simeq 550\,{\rm MeV}$ and
$m_V = (m_{\rho} + m_{m_\omega})/2 \simeq 775\,{\rm MeV}$,
we arrive to the ratios:
\begin{eqnarray}
  \frac{C_b^{\rm sat}}{C_a^{\rm sat}} {\Big|}_{\Lambda \sim (m_{S}-m_{V})} &\sim& 0.20 \, ,
  \label{eq:sat-Cb} \\
  \frac{D_a^{\rm sat}}{C_a^{\rm sat}} {\Big|}_{\Lambda \sim (m_{S}-m_{V})} &\sim& 1.00 \, ,
  \label{eq:sat-Da} \\
  \frac{E_b^{\rm sat}}{D_a^{\rm sat}} {\Big|}_{\Lambda \sim (m_{S}-m_{V})} &\sim& 0.20 \, .
  \label{eq:sat-Eb}
\end{eqnarray}
The first one of these ratios was checked in Ref.~\cite{Liu:2019zvb},
where it was respected at the $30\%$ level.
If we determine $C_a$ and $C_b$ as in Ref.~\cite{Liu:2019tjn}, we obtain
$| C_b / C_a | = 0.16$ for $\Lambda = 0.75\,{\rm GeV}$, which reproduces
the absolute magnitude of Eq.~(\ref{eq:sat-Cb}) at the $25\%$ level.
However we note that in Ref.~\cite{Liu:2019tjn} the sign of $C_b$
depends on which are the spin of the $P_c(4440)$ and $P_c(4457)$
pentaquarks: the sign is correctly reproduced if
the $P_c(4457)$ is $J=\tfrac{1}{2}$.
The second one ($C_a = D_a$) appears for instance in Ref.~\cite{Wu:2010jy},
which predicted the $P_{cs}$, and
Refs.~\cite{Wu:2010vk,Xiao:2019gjd,Shen:2019evi} as a consequence of
the universality of the vector meson coupling.
The recent work of Ref.~\cite{Wang:2019nvm} uses the chiral quark model
to guess the contact-range couplings, leading to $C_a \simeq D_a$ too.
The third relation has not been previously used, as $E_b$ is usually
set to zero (e.g. Ref.~\cite{Xiao:2019gjd}, where their $E_b$ equivalent
is called $\hat{\mu}_{24}$).
Finally it is worth noticing the following
\begin{eqnarray}
  \frac{C_b^{\rm sat}}{C_a^{\rm sat}} \sim \frac{E_b^{\rm sat}}{D_a^{\rm sat}} \, ,
\end{eqnarray}
where this relation actually does not depend so much on saturation being
correct or accurate, but rather on the fact that the light-meson
exchange potentials are identical under
the approximations we have made.
In the following lines we will explore the consequences of
this relation.



\section{Accidental symmetry in the pentaquark potential}
\label{sec:accidental}

Light-meson exchange actually suggests a very interesting relation between
the $P_c(4440)$, $P_c(4457)$ and the two spin states of $\bar{D}^* \Xi_c$.
The S-wave light-meson exchange potential in the $\bar{H}_c S_c$ system is
\begin{eqnarray}
  V_{\rm OBE}(\bar{H}_c S_c) = V_a + V_b \, \vec{\sigma}_L \cdot \vec{S}_L \, ,
\end{eqnarray}
which for the $J=\tfrac{1}{2}$, $\tfrac{3}{2}$ $\bar{D}^* \Sigma_c$ systems
reads
\begin{eqnarray}
  V_{\rm OBE}(\bar{D}^* \Sigma_c, \tfrac{1}{2})
  &=& V_a - \frac{4}{3} V_b \, ,  \\
  V_{\rm OBE}(\bar{D}^* \Sigma_c, \tfrac{3}{2})
  &=& V_a + \frac{2}{3} V_b \, ,  
\end{eqnarray}
from which we expect the hyperfine splitting to be proportional to
\begin{eqnarray}
  M(\bar{D}^* \Sigma_c, \tfrac{3}{2}) - M(\bar{D}^* \Sigma_c, \tfrac{1}{2})
  \propto 2 V_b \, .
\end{eqnarray}
In comparison for the coupled $\bar{H}_c T_c$-$\bar{H}_c S_c$ system,
the corresponding potential reads
\begin{eqnarray}
  V_{\rm OBE}(\bar{H}_c T_c,\bar{H}_c S_c) =
  \begin{pmatrix}
    W_a  & W_b \, \vec{\sigma}_L \cdot \vec{\epsilon}_L \\
    W_b \, \vec{\sigma}_L \cdot \vec{\epsilon}_L &
    V_a + V_b \, \vec{\sigma}_L \cdot \vec{S}_L \\
    \end{pmatrix}
\end{eqnarray}
where if we consider vector meson exchange, vector meson dominance,
the quark model relations for the charmed baryon magnetic moments
and SU(3) symmetric sigma exchange, we will have
\begin{eqnarray}
  V_a \simeq W_a \qquad \mbox{and} \quad V_b \simeq W_b \, .
  \label{eq:accidental}
\end{eqnarray}
This will receive small corrections from $\eta$-exchange (that work in the
direction of making $| W_b | > | V_b|$), which we will ignore
as they are small.
Now, in the limit where the $\bar{D} \Xi_c'$-$\bar{D} \Xi_c$ and
$\bar{D} \Xi_c^*$-$\bar{D}^* \Xi_c$ thresholds are degenerate
and have the same mass, there will be two eigenvalues
for this potential, which correspond to the linear combinations
\begin{eqnarray}
  | \bar{D}^* \Xi_c (J=\tfrac{1}{2}) (\pm) \rangle &=&
  \frac{1}{\sqrt{2}}\left[ |\bar{D} \Xi_c' \rangle \pm
    |\bar{D} \Xi_c^* \rangle \right] \, , \\
  | \bar{D}^* \Xi_c (J=\tfrac{3}{2}) (\pm) \rangle &=&
  \frac{1}{\sqrt{2}}\left[ |\bar{D}^* \Xi_c \rangle \pm
    |\bar{D} \Xi_c^* \rangle \right] \, ,
\end{eqnarray}
with potential eigenvalues
\begin{eqnarray}
  V_{\rm OBE}(J=\tfrac{1}{2},\tfrac{3}{2}, \pm) = \frac{1}{2} (W_a + V_a) \pm W_b
  \simeq V_a \pm V_b \, . \label{eq:Pcs-split-OBE}
\end{eqnarray}
In particular, in this limit the hyperfine splitting between
the ``$\pm$ states'' will be
\begin{eqnarray}
  M(\bar{D}^* \Xi_c, +) - M(\bar{D}^* \Xi_c, -)
  \propto 2 W_b \simeq 2 V_b \, , \label{eq:Pcs-sign}
\end{eqnarray}
that is, expected to be similar to the hyperfine splitting between
the $J=\tfrac{1}{2}$ and $\tfrac{3}{2}$ $\bar{D}^* \Sigma_c$ states.

Of course, this accidental symmetry in the potential will be broken by the fact
that there is a mass gap between the involved coupled channels.
The effect of this mass gap will be to decrease the hyperfine splitting
and to force the ``$\pm$'' sign in Eq.~(\ref{eq:Pcs-split-OBE})
as to make the state corresponding to the lower threshold
the most attractive configuration.
If we compare the characteristic momentum scale of the coupled channel
exchange potential (i.e. the vector meson mass $m_{V}$) and
the momentum scale of the coupled channel dynamics
($\Lambda_{CC} = \sqrt{2 \mu \Delta_{CC}} \simeq 266$
and $279\,{\rm MeV}$), the ratio is $0.34$ and $0.36$
for the $\bar{D} \Xi_c'$ and $\bar{D} \Xi_c^*$ channels,
respectively.
Together they add to $0.70$, i.e. we expect the hyperfine splittings to be
about $30\%$ of the expected value were not to be a mass gap.
For the $J=\tfrac{1}{2}$ ($\tfrac{3}{2}$) configuration,
the $\bar{D}^* \Xi_c$ threshold is heavier (lighter) than
the $\bar{D} \Xi_c'$ ($\bar{D} \Xi_c^*$) one,
which forces the most repulsive (attractive) sign configuration\
in Eq.~(\ref{eq:Pcs-sign}).
\begin{eqnarray}
  M(\bar{D}^* \Xi_c, \tfrac{1}{2}) - M(\bar{D}^* \Xi_c, \tfrac{3}{2})
  &\propto& 2 |W_b| \left( 1 - \sum_{CC}\mathcal{O}(\frac{\Lambda_{CC}}{m}) \right)
  \nonumber \\
  &\simeq& 2 |V_b| \left( 1 - \sum_{CC}\mathcal{O}(\frac{\Lambda_{CC}}{m}) \right) \, .
\end{eqnarray}
As a consequence, if the previous approximations hold, the hyperfine splitting
of the two $\bar{D}^* \Xi_c$ pentaquarks will be similar to the one of
the two $\bar{D}^* \Sigma_c$ pentaquarks, i.e.
\begin{eqnarray}
  | M(\bar{D}^* \Xi_c, \tfrac{1}{2}) - M(\bar{D}^* \Xi_c, \tfrac{3}{2}) |
  \simeq
  17\,{\rm MeV} \, .
\end{eqnarray}
However if the effect is to be reduced by a $70\%$, as suggested by the scale
comparison, we will end up with a $5.1\,{\rm MeV}$ splitting.
Though the sign of the hyperfine splitting might be protected owing to power
counting and the nature of the coupled channel dynamics, its size will be
diminished owing to the finite mass gaps between the channels.
Besides, the uncertainties in the couplings of the light-meson exchange
potential are also large.
In the following we will see how this expected effect holds when compared
with different error sources.

For doing the explicit calculations we first obtain the $C_a$ and $C_b$
couplings from the masses of the $P_c(4440)$ and $P_c(4457)$ pentaquarks,
i.e. the calculation of Ref.~\cite{Liu:2019tjn}, where we notice that
for the hyperfine splitting it does not matter which spin is
each pentaquark~\footnote{
  This merely changes the sign of $C_b$, which is later identified with $E_b$,
  but the coupled channel effects do not depend on the sign of the later
  coupling.},
resulting in $C_a = -2.52\,(-0.85)\,{\rm fm}^2$,
$C_b = \pm 0.54\,(0.11)\,{\rm fm}^2$ for
$\Lambda = 0.5\,(1.0)\,{\rm GeV}$.
Then we consider the phenomenological potential symmetry of
Eq.~(\ref{eq:accidental}) as applied to the contact-range couplings
(i.e. $C_a = D_a$, $C_b = E_b$), from which we get:
\begin{eqnarray}
  M(P_{cs},\tfrac{1}{2}) &=& 4464.5 - 1.1 i\,(4464.4 - 1.3 i)\,{\rm MeV} \, ,
  \label{eq:MPcs-12-accidental} \\
  M(P_{cs},\tfrac{3}{2}) &=& 4459.9 (4459.7)\,{\rm MeV} \, ,
\end{eqnarray}
for $\Lambda = 0.5\,(1.0)\,{\rm GeV}$, with the hyperfine splitting
\begin{eqnarray}
  \Delta M(P_{cs}) = 4.6\,(4.7) \,{\rm MeV} \, , \label{eq:split-sym}
\end{eqnarray}
which indeed indicates a reduction of the coupled channel effects owing to the
finite mass gap between the channels, and where from now on we define
\begin{eqnarray}
  \Delta M(P_{cs}) = M(\tfrac{1}{2}) - M(\tfrac{3}{2}) \, .
\end{eqnarray}
Yet, even if this accidental symmetry is greatly reduced in the hyperfine
splittings, it is worth noticing that the predictions we obtain
from using the effective $C_a$ and $C_b$ couplings describing
the $P_c(4440)$ and $P_c(4457)$ pentaquarks are basically
compatible with the experimental mass of
the observed $P_{cs}$ pentaquark.



\section{The hyperfine splitting}
\label{sec:hyperfine}

\begin{table*}[t]
  \begin{center}
\begin{tabular}{|lccc|}
\hline \hline
Assumptions: & $J$ & $M(J)$ & $M(J=1/2) - M(J=3/2)$ \\
\hline 
\hline 
EFT A: $\,\, {E_b}/{D_a} \sim {Q}/{M}$ and & ${1}/{2}$ & $4458.8$ (Input) &
    {\multirow{2}{*}{$21.5\,(11.6-35.4)$}} \\
    $D_a$ from $P_{cs}(4459)$ as $J=\frac{1}{2}$ & ${3}/{2}$ & $4437.3\,(4423.3-4447.2)$ &  \\
    \hline
EFT B: $\,\, {E_b}/{D_a} \sim {Q}/{M}$ and & ${1}/{2}$ & $4469.5\,(4465.7-4473.4)$ &
        {\multirow{2}{*}{$10.7\,(6.9-14.6)$}} \\
        $D_a$ from $P_{cs}(4459)$ as $J=\frac{3}{2}$ & ${3}/{2}$ & $4458.8$ (Input) &  \\
        \hline 
        \hline 
EFT+RG A: $\,\, {E_b}/{D_a} \sim \mathcal{F}(Q/\Lambda)\,{Q}/{M}$ & ${1}/{2}$ & $4458.8$ (Input) &
    {\multirow{2}{*}{$30.6\,(27.4-40.4)$}} \\
    and $D_a$ from $P_{cs}(4459)$ as $J=\frac{1}{2}$ & ${3}/{2}$ & $4428.1\,(4417.7-4431.5)$ &  \\
    \hline
EFT+RG B: $\,\, {E_b}/{D_a} \sim \mathcal{F}(Q/\Lambda)\,{Q}/{M}$ & ${1}/{2}$ & $4471.4\,(4470.7-4471.7)$ &
        {\multirow{2}{*}{$12.6\,(11.9-12.9)$}} \\
        and $D_a$ from $P_{cs}(4459)$ as $J=\frac{3}{2}$ & ${3}/{2}$ & $4458.8$ (Input) &  \\
        \hline \hline 
        Accidental: $\,\, C_a \sim D_a$, $C_b \sim E_b$ & ${1}/{2}$ & $4464.5\,(4464.40-4464.5)$ &
    {\multirow{2}{*}{$4.7\,(4.6-4.7)$}} \\
    and $C_a$, $C_b$ from $P_{c}(4440/4457)$ & ${3}/{2}$ & $4459.8\,(4459.7-4459.9)$ &  \\
    \hline \hline 
    Saturation I: $\,\, C_a \sim D_a$ and
    & ${1}/{2}$ & $4473.1\,(4471.1-4475.3)$ &
    {\multirow{2}{*}{$14.3\,(12.7-16.5)$}} \\
    $E_b$ from $P_{cs}(4459)$ as $J=\tfrac{3}{2}$ & ${3}/{2}$ & $4458.8$ (Input) &  \\
    \hline
    Saturation II:   & ${1}/{2}$ & $4469.6\,(4469.2-4470.1)$ &
    {\multirow{2}{*}{$5.3\,(3.4-8.8)$}} \\
    $C_a \sim D_a$ and $E_b \sim 0.2 D_a$ & ${3}/{2}$ & $4464.3\,(4461.3-4466.6)$ &  \\
    \hline \hline
    Two-peak solution: & ${1}/{2}$ & $4467.8$ (Input) &
    {\multirow{2}{*}{$12.9$ (Input)}} \\
    $C_a = 0.90 D_a$ and $E_b = 0.28 D_a$ & ${3}/{2}$ & $4454.9$ (Input) &  \\
    \hline
    \hline
\end{tabular}
\caption{Expected masses and hyperfine splitting (in units of ${\rm MeV}$) of
  the $J=\tfrac{1}{2}$ $\bar{D} \Xi_c'$-$\bar{D}^* \Xi_c$ and
  $J=\tfrac{3}{2}$ $\bar{D}^* \Xi_c$-$\bar{D} \Xi_c^*$ pentaquarks
  with the different assumptions considered in this work.
  They include power counting estimations from EFT (``EFT A \& B''),
  the RG-improved EFT (``EFT+RGA A\&B'', where RG stands
  for ``renormalization group''),
  the accidental symmetry of the scalar and vector meson exchange
  potentials between the $P_c(4440/4457)$ and the new $P_{cs}(4459)$
  (``Accidental''), and the saturation relations
  Eqs.~(\ref{eq:sat-Da}) and (\ref{eq:sat-Eb})
  (``Saturation I \& II'').
  The central values correspond to $\Lambda = 0.75$ ${\rm GeV}$,
  while the spread to $\Lambda = 0.5-1.0$ ${\rm GeV}$.
  At the end (``Two-peak solution'') we compare with the two-peak
  fit of Ref.~\cite{Aaij:2020gdg}, from which an hyperfine
  splitting of $12.9 \pm 4.6\,{\rm MeV}$ is expected.
  If this figure turns out to be confirmed by future experimental studies,
  this will discard the ``EFT A'', ``EFT+RGA A'', ``Accidental'' and
  ``Saturation II'' estimations.
  } \label{tab:Pcs-splitting}
\end{center}
\end{table*}

Now we will analyze the possible size of the hyperfine splittings
with the (admittedly approximate) information we have derived
from light-meson exchange saturation.
We begin with the simplest of the relations, that is:
\begin{eqnarray}
  \frac{D_a}{C_a} \sim 1 \, ,
\end{eqnarray}
and determine $C_a$ from reproducing the $P_c(4312)$ pentaquark, which yields
$C_a = -1.19$ ($-(2.17-0.80)$) ${\rm fm}^2$ for $\Lambda = 0.75$ ($0.5-1.0$)
$\rm GeV$, where we will use a central value of the cutoff close to
the expected scale at which saturation works (i.e. $\Lambda \sim m_V$),
while we will still consider the $0.5-1.0\,{\rm GeV}$ cutoff range
for estimating the regulator uncertainties.
If we set $E_b = 0$, it will lead to
\begin{eqnarray}
  M(P_{cs}, J = \tfrac{1}{2},\tfrac{3}{2}) = 4467.5\,(4466.9-4468.1)\,{\rm MeV}
  \, ,
\end{eqnarray}
that is, we obtain two degenerate $P_{cs}$ pentaquarks.
If we allow for $E_b \neq 0$ we can effectively fit one of the pentaquarks
to the experimental mass. We obtain~\footnote{Notice that
  the $J=\tfrac{1}{2}$ state will acquire a small finite
  width as it can decay into $\bar{D}\Xi_c'$.
  For convenience we will ignore this width from now on, as it
  is usually of the order of a few ${\rm MeV}$,
  see Eq.~(\ref{eq:MPcs-12-accidental}), and not representative of the full
  width of the $P_{cs}$, which comprises more decay channels.}.:
\begin{eqnarray}
  M(P_{cs},\tfrac{1}{2}) &=& 4473.1 (4471.1-4475.3)\,{\rm MeV} \, , \\
  M(P_{cs},\tfrac{3}{2}) &=& 4458.8\,{\rm MeV} \, ,
\end{eqnarray}
where it should be noticed that scenario B is automatically chosen,
as for $C_a / D_a \sim 1$ the mass of the $P_{cs}$ pentaquark
can only be reproduced if $J=\tfrac{3}{2}$.
The hyperfine splitting will be
\begin{eqnarray}
  \Delta M(P_{cs}) = 14.3\,(12.7-16.5)\,{\rm MeV} \, ,
\end{eqnarray}
which is definitely larger than the estimation from the phenomenological
symmetry in the pentaquark potential.
The ratio $E_b/D_b = 0.34\,(0.24-0.53)$ will also be larger than the expectation
from saturation, Eq.~(\ref{eq:sat-Eb}).

Alternatively we can assume that the saturation relations
in Eqs.~(\ref{eq:sat-Da}) and (\ref{eq:sat-Eb}) both hold,
in which case the masses of the two $P_{cs}$'s are
\begin{eqnarray}
  M(P_{cs},\tfrac{1}{2}) &=& 4469.6 \,(4469.2-4470.1)\, {\rm MeV} \, , \\
  M(P_{cs},\tfrac{3}{2}) &=& 4464.3 \,(4461.3-4466.6)\, {\rm MeV} \, ,
\end{eqnarray}
and the hyperfine splitting is
\begin{eqnarray}
  \Delta M (P_{cs}) = 5.3\,(3.4-8.8) \, {\rm MeV} \, ,
\end{eqnarray}
which is closer to the one we derived from the phenomenological symmetry,
i.e. Eq.~(\ref{eq:split-sym}).

Now there are a series of (potentially large) uncertainties related to
the previous relations.
The most obvious one is the existence of SU(3)-breaking corrections
between the coupling
in the $\bar{D} \Sigma_c$ (strangeness $S=0$) and the
$\bar{D} \Xi_c'$, $\bar{D} \Xi_c^*$ (strangeness $S=-1$) systems,
i.e. between the $P_c(4312)$ and $P_{cs}'$, $P_{cs}^*$ pentaquarks.
Their couplings, which are identical if SU(3)-flavor symmetry is exactly
preserved, will differ by a correction
\begin{eqnarray}
  C_a(P_{cs}^{('/*)}) = C_a(P_c) + \delta C_a^{\slashed{F}} \, .
\end{eqnarray}
where $\delta C_a^{\slashed{F}}$ indicates the correction to the $C_a$ coupling.
From a comparison with the pion and kaon weak decay constants,
$f_{\pi} \simeq 130\,{\rm MeV}$ and $f_{K} \simeq 160\,{\rm MeV}$,
we expect the size of the SU(3)-breaking effects
to be of the order of
\begin{eqnarray}
  \frac{\delta C_a^{\slashed{F}}}{C_a} \sim
  \left( \frac{f_K - f_{\pi}}{f_{\pi}} \right) \sim 0.23 \, .
\end{eqnarray}
Alternatively, we can consider this problem from the point of view of chiral
symmetry~\footnote{Notice that the EFT we are using here is not a pionless EFT,
  but rather a pionful EFT for which pion (and pseudoscalar meson) exchanges
  are considered to be subleading and thus not explicitly included
  in the leading order description. Thus chiral symmetry
  considerations can play an explicit role.},
in which the $C_a$ coupling can be decomposed into quark-mass
independent and quark-mass dependent pieces~\cite{Petschauer:2020urh}:
the quark-mass dependent piece comes from terms in the Lagrangian
where the quark-mass matrix is inserted between the hadron fields,
with these terms expected to be subleading (as in this case
we are expanding around the massless quark limit).
This quark-mass dependence can be translated into a quadratic dependence
on the mass of the pseudoscalar Nambu-Goldstone bosons, which can be
schematically written as
\begin{eqnarray}
  C_a(P_{c}) &=&
  C_a^{[0,0]} + C_a^{[1,0]}\,\frac{m_{\pi}^2}{\Lambda_{\chi}^2} + C_a^{[0,1]}\,
  \frac{m_{\pi}^2}{\Lambda_{\chi}^2} + \dots
  \, , \\
  C_a(P_{cs}^{('/*)}) &=&
  C_a^{[0,0]} + C_a^{[1,0]}\,\frac{m_{\pi}^2}{\Lambda_{\chi}^2} + C_a^{[0,1]}\,
  \frac{m_{K}^2}{\Lambda_{\chi}^2} + \dots
  \, , 
\end{eqnarray}
where $C^{[n,m]}$ indicates $n$ ($m$) insertions of the quark mass matrix
between the charmed antimeson (baryon) fields, $m_{\pi}$ and $m_K$ are
the pion and kaon masses and $\Lambda_{\chi} \sim 1\,{\rm GeV}$ is
the chiral symmetry breaking scale.
Rearranging the terms, the difference can be rewritten in terms of
a chiral symmetry breaking ($\chi SB$) correction, which takes the form
\begin{eqnarray}
  C_a(P_{cs}^{('/*)}) = C_a(P_c) + \delta C_a^{\chi SB} \, .
\end{eqnarray}
From the arguments about quark-mass dependence shown above,
the size of this correction is expected to be of the order
\begin{eqnarray}
  \frac{\delta C_a^{\chi SB}}{C_a} \sim
  \left( \frac{m_K^2 - m_{\pi}^2}{\Lambda_{\chi}^2} \right)
  \sim 0.23 \, ,
\end{eqnarray}
where it is interesting to notice that its size is identical to
the previous estimation based on $f_{\pi}$ and $f_{K}$.
The experience in the light-baryon sector suggests breakings larger
than the previous estimation~\cite{Haidenbauer:2014rna},
where repulsion increases with the number of strange quarks.
The recent discovery of the $Z_{cs}$~\cite{Ablikim:2020hsk}
allows for a comparison of the couplings
required to reproduce the $Z_c$/$Z_c^*$~\cite{Albaladejo:2015lob} and
$Z_{cs}$~\cite{Yang:2020nrt} as virtual states,
with attraction apparently increasing with the strange quark content
(though the uncertainties are large).

Independently of the derivation, if SU(3)-breaking corrections
reach a certain level they would lead to unbound $\bar{D} \Xi_c'$
and $\bar{D} \Xi_c^*$ systems.
In particular this will happen for
\begin{eqnarray}
  | C_a(P_{cs}') | \leq 0.65\, (0.52-0.70) \, | C_a (P_c) | \, ,
\end{eqnarray}
for $\Lambda = 0.75 (0.5-1.0)\,{\rm GeV}$, where we indicate
which pentaquark we are referring to in the parentheses.
In this regard the previous $20\%$ estimations indicate that
the SU(3)-symmetry partner of the $P_c(4312)$ pentaquark is
still likely to bind.
The eventual observation of a $\bar{D} \Xi_c'$ molecular pentaquark
could bring light to this issue.

A significant effect which might influence the size of SU(3)-flavor
breaking corrections is the nature of sigma meson exchange:
if the sigma were not to couple with the strange meson,
saturation will suggest
\begin{eqnarray}
  \frac{C_a^{\rm sat}(P_{cs}')}{C_a^{\rm sat}(P_c)} \sim 0.62 \, ,
  \label{eq:ratio-std-sigma}
\end{eqnarray}
which is in the limit between binding and not binding for $P_{cs}'$
(though in the no-binding case, the $P_{cs}'$ pentaquark
might still survive as a virtual state).
Additionally, in this scenario from saturation we will expect 
\begin{eqnarray}
  \frac{D_a}{C_a(P_{cs}')} \sim 1.0 \quad \mbox{and} \quad
  \frac{E_b}{D_a} \sim 0.33 \, ,
\end{eqnarray}
which would imply that the $P_{cs}$ could also very well be close to not
binding (except for the increase in the relative strength of $E_b$, which
would help in the $J=\tfrac{3}{2}$ configuration).
With these coupling ratios, the masses of the molecular $P_{cs}$ pentaquarks
would be
\begin{eqnarray}
  M(P_{cs},\tfrac{1}{2}) &=& 4478.0\,(4476.4-4478.1)\,{\rm MeV} \, , \\
  M(P_{cs},\tfrac{3}{2}) &=& 4477.7\,(4476.7-4478.0)\,{\rm MeV} \, ,
\end{eqnarray}
with the $J=\tfrac{1}{2}$ state in the second Riemann sheet with respect to
the $D^* \Xi_c$ threshold (i.e. it is a shallow virtual state / resonance).
That is, only the $J=\tfrac{3}{2}$ state is an actual bound state.
In this later case the hyperfine splitting is compatible with zero and
can even change signs as the $J=\tfrac{1}{2}$ is allowed
to be a virtual state.
However this large SU(3)-breaking in the direction of making
the $\bar{D}^* \Xi_c$ less bound does not reproduce
the experimental mass of the $P_{cs}$ pentaquark
(unless we allow for $E_b/D_a \sim 1.23$, which is a considerable deviation
from the saturation relations and would  lead to a hyperfine
splitting of $50.2$ $\rm MeV$).
Thus it is unlikely that SU(3)-breaking would be
as large as in Eq.~(\ref{eq:ratio-std-sigma}), at least if
its effect is to reduce attraction in the $\bar{D} \Xi'$ system.

Finally, it is possible to make a comparison with the two-peak fit included
in Ref.~\cite{Aaij:2020gdg}, which leads to two $P_{cs}$ pentaquarks
with masses $4454.9 \pm 2.7\,{\rm MeV}$ and $4467.8 \pm 3.7\,{\rm MeV}$,
respectively.
Concrete calculations assuming that the lighter (heavier) $P_{cs}$ is
a $J=\tfrac{3}{2}$ ($\tfrac{1}{2}$) $\bar{D}^* \Sigma_c$ molecule yield
$D_a = -1.31$ $(-(2.59-0.86))\,{\rm fm^2}$ and
$|E_b| = 0.36$ $(0.95-0.19)\,{\rm fm}^2$ for
$\Lambda = 0.75 (0.5-1.0)\,{\rm GeV}$.
This determination of the couplings, together with the previous
determination of $C_a$ from the $P_c(4312)$, provide the ratios
\begin{eqnarray}
  \frac{D_a}{C_a} &=& 1.11\,(1.19-1.07) \, , \label{eq:rat-1} \\
  \frac{E_b}{D_a} &=& 0.28\,(0.37-0.22) \, , \label{eq:rat-2}
\end{eqnarray}
where the central value and the spreads correspond to
$\Lambda = 0.75\,(0.5-1.0)\,{\rm GeV}$, as usual.
If the two-peak fit ends up being confirmed in future studies,
the previous indicates a bit more attraction than expected
for the $D_a$ coupling
(but compatible with the errors we would expect for a phenomenological
determination of the $D_a/C_a$ ratio) and that the size of $E_b$ seems
to be underestimated by the phenomenological arguments we provide
(yet compatible with the power counting estimates we proposed).

We summarize the different estimations we have considered along this work
in Table \ref{tab:Pcs-splitting}.
These indicate that, though it is not possible to determine
the hyperfine splitting accurately from theory alone,
power counting arguments and phenomenological approximations
suggest it might be in the $5-15\,{\rm MeV}$ range
(i.e. compatible with the two-peak fit in~\cite{Aaij:2020gdg}),
with the $J=3/2$ pentaquark being the lighter state.
For comparison, Refs.~\cite{Wu:2010jy,Wu:2010vk} predict degenerate
$P_{cs}$ pentaquarks.
Meanwhile, a recent calculation in the one-boson-exchange model generates a
$\Delta M(P_{cs}) = (2.4-20.0)\,{\rm MeV}$ splitting, which also comes from the
$\bar{D} \Xi_c'$-$\bar{D}^* \Xi_c$ and $\bar{D}^* \Xi_c$-$\bar{D} \Xi_c^*$
coupled channel dynamics~\cite{Zhu:2021lhd}.
In contrast Ref.~\cite{Wang:2019nvm} obtains
$\Delta M(P_{cs}) = -6.0\,{\rm MeV}$ from two-pion-exchange (TPE).
This is interesting as the naive expectation would be that TPE is of order
$(Q/M)^2$ in the EFT expansion and we might expect it to play a minor role.
Thus, the role of TPE might indeed deserve further attention in the future.
However there is the practical limitation that this calculation
will also require $(Q/M)^2$ corrections to the contact-range potential,
i.e. more unknown parameters.

Regarding which is the spin of the $P_{cs}$ pentaquark,
from the different predictions in Table \ref{tab:Pcs-splitting} it seems
that the $J=\tfrac{3}{2}$ $\bar{D} \Xi_c^*$ configuration might provide
a better match to its experimental mass, though uncertainties are
too large to draw definite conclusions.
It is also important to stress that the experimental determination of
resonance masses usually relies in using the Breit-Wigner parametrization.
Other parametrizations might yield different masses,
as happened with the $P_c(4312)$~\cite{Fernandez-Ramirez:2019koa},
the $Z_c(3900)$ / $Z_c(4020)$~\cite{Albaladejo:2015lob},
and the $Z_{cs}(3985)$~\cite{Yang:2020nrt}. 
Thus the mass of a molecular $P_{cs}$ might not coincide
with the experimental determination of Ref.~~\cite{Aaij:2020gdg}.


\section{Decay into $J/\psi \Lambda$}
\label{sec:decays}

It is interesting to notice that the observation of the $P_{cs}(4459)$
in the $J/\psi \Lambda$ channel might provide further
circumstantial evidence of its spin.
If we decompose the $J = \tfrac{1}{2}$, $\tfrac{3}{2}$ $\bar{D}^* \Xi_c$ system
into its heavy- and light-quark spin components, we find
\begin{eqnarray}
  | \bar{D}^* \Xi_c (J=\tfrac{1}{2}) \rangle &=&
  \left( \frac{\sqrt{3}}{2} 0_H + \frac{1}{2} 1_H \right)
  \otimes {\frac{1}{2}}_L \, , \\
  | \bar{D}^* \Xi_c (J=\tfrac{3}{2}) \rangle &=&
  1_H \otimes {\frac{1}{2}}_L \, ,
\end{eqnarray}
which is to be compared with
\begin{eqnarray}
  | J/\psi \Lambda \rangle = 1_H \otimes {\frac{1}{2}}_L \, ,
\end{eqnarray}
where $S_H$ and $S_L$ refer to the heavy- and light-quarks spin.
If the decay preserves HQSS~\cite{Sakai:2019qph}
the expect the following relation between matrix elements:
\begin{eqnarray}
  \langle J/\psi \Lambda | H | \bar{D}^* \Xi_c (J=\tfrac{1}{2}) \rangle =
  \frac{1}{2}\,
  \langle J/\psi \Lambda | H | \bar{D}^* \Xi_c (J=\tfrac{3}{2}) \rangle \, ,
  \nonumber \\
\end{eqnarray}
which for degenerate $\bar{D}^* \Xi_c$ states implies that the partial decay
widths of the $J = \tfrac{1}{2}$ and $\tfrac{3}{2}$ configurations will
show a $1\, : \, 4$ ratio.
In fact phenomenological calculations seems to support these ratios,
with Ref.~\cite{Xiao:2019gjd} giving a $1\,:\,1.78$ ratio
for the amplitudes / couplings and Ref.~\cite{Xiao:2021rgp}
yielding $1\,:\,4.35$ for the partial decay widths.
Of course, this does not determine the spin of the $P_{cs}(4459)$,
but nonetheless indicates that, {\it ceteris paribus}, the probability of
discovering the $J = \tfrac{3}{2}$ molecule in the $J/\psi \Lambda$
invariant mass distribution might be larger than
for its $J = \tfrac{1}{2}$ partner.
But this conclusion is dependent on the production rates
from the $\Xi_b$ decays, which have been recently investigated
in Ref.~\cite{Wu:2021dmq}, suggesting that the production rate of a
$J=\tfrac{3}{2}$ $\bar{D}^* \Xi_c$ pentaquark would be 4.9 times
the one for its $J=\tfrac{1}{2}$ partner.
Ideally, it would be possible to adapt the methods
of Refs.~\cite{Du:2019pij,Du:2021fmf} (originally formulated for
the three $P_c$ pentaquarks) to analyze the invariant mass distribution
data of the new $P_{cs}$.



\section{Conclusions}
\label{sec:conclusions}

The $P_{cs}(4459)$ is the latest piece of the pentaquark puzzle.
Its closeness to the $\bar{D}^* \Xi_c$ threshold suggests that
it might be a bound state of these hadrons.
Then the question is what is the connection of the $P_{cs}(4459)$
with the well-known $P_c(4312)$, $P_c(4440)$
and $P_c(4457)$ in the molecular hypothesis.
At first sight the answer is unclear:
the $\bar{D}^* \Xi_c$ system is not directly connected by SU(3)-flavor
and HQSS with the $\bar{D} \Sigma_c$ and $\bar{D} \Sigma_c^*$ systems,
which are the usual molecular interpretations of the three $P_c$ pentaquarks.
However if we resort to phenomenological arguments then we can bridge
the gap between the new $P_{cs}$ and the previous $P_c$'s, resulting
in a coherent description of these four pentaquarks.
From vector meson dominance and the quark model, we point out a possible
accidental symmetry between the potentials in the $P_c$ and $P_{cs}$
sectors, though owing to its phenomenological nature large
deviations are to be expected.

There are two possible spin configurations for a molecular $P_{cs}$,
which in principle should be degenerate.
In this regard the explicit inclusion of
the $\bar{D} \Xi_c'$-$\bar{D}^* \Xi_c$ and $\bar{D}^* \Xi_c$-$\bar{D} \Xi_c^*$
coupled channel dynamics, which is required by power counting arguments,
breaks this degeneracy and thus might have important implications
for spectroscopy.
This mechanism generates a sizable hyperfine splitting which we estimate
to be in the $5-15\,{\rm MeV}$ range.
Incidentally, this estimation is in line with the proposed two-peak
solution in~\cite{Aaij:2020gdg}.

In general the predicted mass of the $J=\tfrac{3}{2}$ $P_{cs}$ pentaquark
are closer to its experimental value than its $J=\tfrac{1}{2}$ partner,
which might be interpreted as favoring the former spin assignment.
But theoretical errors in the masses make it unpractical to determine
the spin of the new pentaquark from spectroscopy alone.
In this regard, the partial decay widths of the two spin configurations
to $J/\psi \Lambda$, where the $P_{cs}$ have been discovered,
approximately differ by a factor of four, making
the $J = \tfrac{3}{2}$ configuration considerably
easier to detect in this channel.
Of course this is only true provided all other effects are similar.
Independently of their spin,
the existence of two possible $\bar{D}^* \Xi_c$ molecules tends to hold up well
within the expected uncertainties of a phenomenological approach.
As happened with the original $P_{c}(4450)$, future experiments could
further determine whether there are really two $\bar{D}^* \Xi_c$
states and which is their mass difference.

\section*{Acknowledgments}
We would like to thank Li-Sheng Geng and Eulogio Oset
for comments on this manuscript and Jorge Segovia
for discussions.
M.P.V. thanks the IJCLab of Orsay, where part of this work has been done,
for its long-term hospitality.
This work is partly supported by the National Natural Science Foundation of
China under Grants No. 11735003, No. 11975041, the fundamental Research Funds
for the Central Universities and the Thousand
Talents Plan for Young Professionals.

\appendix

\section{Power counting arguments with running coupling constants}
\label{sec:running}

In this work we have made use of power counting estimations of the size of
the contact-range couplings appearing in the EFT description of
the molecular pentaquarks.
However the contact-range couplings are cutoff dependent and the aforementioned
estimations were originally formulated for the {\it renormalized couplings}~\cite{vanKolck:1998bw},
the definition of which is scheme dependent.
Yet, it has been argued that these estimations do indeed apply
to running couplings if the cutoff is sufficiently
soft~\cite{Valderrama:2016koj,Epelbaum:2017byx}.
But the previous arguments are qualitative in nature:
here we will see how to modify dimensional estimations
in a quantitative manner for their use with running couplings.
This will be particularly useful for the dimensional estimation of the ratio
of the $D_a$ and $E_b$ coupling constant employed in Sect.~\ref{sec:pc}
(check Eq.~(\ref{eq:coupling-ratio})).

The easiest example will be a contact-range theory with only one channel,
in which we determine the coupling from the condition of reproducing
the two-body scattering amplitude.
For instance, if we consider the scattering of a $\bar{D}^*$ antimeson and
a $\Xi_c$ baryon, the T-matrix can be written as
\begin{eqnarray}
  \langle p' | T(k) | p \rangle =
  f(\frac{p'}{\Lambda})\,
  \frac{1}{\frac{1}{D_a(\Lambda)} - I_0(k, \Lambda)} \,
  f(\frac{p}{\Lambda})\, \, ,
\end{eqnarray}
where $f(x)$ is the regulator we use for the contact-range potential and 
with $I_0(k,\Lambda)$ the loop function:
\begin{eqnarray}
  I_0(k,\Lambda) =
  \int \frac{d^3 q}{(2 \pi)^3} \frac{f^2(\frac{q}{\Lambda})}
       {\frac{k^2}{2 \mu} - \frac{q^2}{2 \mu}} \, ,
\end{eqnarray}
the exact evaluation of which depends on the details of the regulator.
If we want the T-matrix to be (exactly) cutoff independent at a given
reference momentum $k_R$, this will lead to the condition
\begin{eqnarray}
  \frac{1}{D_a(\Lambda_1)} - I_0(k_R, \Lambda_1)  =
  \frac{1}{D_a(\Lambda_2)} - I_0(k_R, \Lambda_2) 
  \, ,
\end{eqnarray}
where $\Lambda_1$ and $\Lambda_2$ are two different cutoffs.
If we choose to renormalize the scattering amplitude at the bound state pole,
i.e. at $k_R = i \gamma_R = i \sqrt{2 \mu B_2}$,
the $D_a$ coupling will be given by
\begin{eqnarray}
  D_a(\Lambda) = \frac{1}{I_0(i \gamma_R, \Lambda)} \, .
\end{eqnarray}
This coupling will exactly reproduce its power counting estimation 
for a privileged cutoff $\Lambda^*$
\begin{eqnarray}
  D_a(\Lambda^*) = -\frac{2 \pi}{\mu \sqrt{2 \mu B_2}} = D_a^R \, ,
\end{eqnarray}
where for a Gaussian regulator $\Lambda^* \simeq 4.3 \sqrt{2 \mu B_2}$,
which for the particular case of the $P_{cs}(4459)$ yields
$\Lambda^* \simeq 886\,{\rm MeV}$.
However, if the cutoff is different from this privileged value,
the dimensional estimations will have to be corrected as follows
\begin{eqnarray}
  \frac{D_a^R}{D_a(\Lambda)} =
  \frac{I_0(i \gamma_R, \Lambda^*)}{I_0(i \gamma_R, \Lambda)} \, .
\end{eqnarray}
It happens that $\Lambda^* \sim 0.9\,{\rm GeV}$, which implies that corrections
will be small for the range of cutoffs considered in this work
(i.e. $\Lambda = 0.5-1.0 \,{\rm GeV}$).
We notice that, in contrast with what is expected in Refs.~\cite{Valderrama:2016koj,Epelbaum:2017byx}, the estimation of $\Lambda^*$ in the $P_{cs}$
pentaquark is rather large and can be hardly considered
to be a soft scale.
However Refs.~\cite{Valderrama:2016koj,Epelbaum:2017byx} deal
with the two-nucleon system and the deuteron happens to be
considerably more shallow than the $P_{cs}$.
Indeed, repeating the previous arguments for the deuteron yields
$\Lambda^* \simeq 196\,{\rm MeV}$, which is of the order of
the pion mass and more in line with the expectations of
Refs.~\cite{Valderrama:2016koj,Epelbaum:2017byx}.

The inclusion of coupled channel effects can be taken into account
by considering the matrix version of the previous renormalization group equation
\begin{eqnarray}
  && {\begin{pmatrix} D_a(\Lambda_1) & E_b(\Lambda_1) \\
    E_b(\Lambda_1) & C_a(\Lambda_1)
  \end{pmatrix}}^{-1}
  -
  \begin{pmatrix}
    I(k_R, \Lambda_1) & 0 \\
    0 & I(k_R', \Lambda_1)
  \end{pmatrix} \nonumber \\
  &=&
   {\begin{pmatrix} D_a(\Lambda_2) & E_b(\Lambda_2) \\
    E_b(\Lambda_2) & C_a(\Lambda_2)
  \end{pmatrix}}^{-1}
  -
  \begin{pmatrix}
    I(k_R, \Lambda_2) & 0 \\
    0 & I(k_R', \Lambda_2)
  \end{pmatrix} \, ,
\end{eqnarray}
which ensures that the T-matrix is cutoff-independent at
the renormalization point $k = k_R$ and
where $k_R'$ refers to the reference
momentum in the second channel.
If we write the previous equations in coefficients, we end up with
\begin{eqnarray}
  \frac{C_a(\Lambda_1)}{det(V_C(\Lambda_1))} - I(k_R, \Lambda_1) &=&
  \frac{C_a(\Lambda_2)}{det(V_C(\Lambda_2))} - I(k_R, \Lambda_2) \, , \\
  \frac{E_b(\Lambda_1)}{det(V_C(\Lambda_1))} &=&
  \frac{E_b(\Lambda_2)}{det(V_C(\Lambda_2))} \, , \\
  \frac{D_a(\Lambda_1)}{det(V_C(\Lambda_1))} - I(k_R', \Lambda_1) &=&
  \frac{D_a(\Lambda_2)}{det(V_C(\Lambda_2))} - I(k_R', \Lambda_2) \, ,
\end{eqnarray}
where $det(V_C) = C_a D_a - E_b^2$ is the determinant of
the contact-range potential.
Combining the last two equations, we end up with
\begin{eqnarray}
  \frac{E_b(\Lambda)}{D_a(\Lambda)} = \frac{1}{1 + \frac{det(V^R)}{D_a^R}\,(I(k_R',\Lambda) - I(k_R',\Lambda^*))}\,\frac{E_b^R}{D_a^R} \, ,
\end{eqnarray}
If for simplicity we assume the same $\Lambda^*$ as in the uncoupled channel
case and that for $\Lambda^*$ all the power counting estimations are followed
as expected (i.e. $| D_a | = \frac{2 \pi}{\mu \gamma_R}$,
$| C_a | = \frac{2 \pi}{\mu \gamma_R'}$ and
$| E_b | = \frac{2 \pi}{\mu M}$, with $\gamma_R$, $\gamma_R'$
the wave number in each of the coupled channels and $M = m_{\rho}$),
we will end up with a correction factor of
\begin{eqnarray}
  \frac{E_b(\Lambda)}{D_a(\Lambda)} = (1.47-0.91)\,\frac{E_b^R}{D_a^R} \, ,
\end{eqnarray}
for $\Lambda = 0.5-1.0\,{\rm GeV}$, which happens to be close to
the original estimation.
However this small change reduces the cutoff dependence of the hyperfine
splitting (which is now a renormalizable quantity), yielding
$27-40$ and $12-13\,{\rm MeV}$ in scenarios $A$ and $B$ respectively
(check the discussion around Eq.~(\ref{eq:PC-running})
in the main text).

\section{Standard and novel saturation procedures in the two-nucleon system}
\label{sec:saturation-NN}

In this appendix we compare the standard and modified saturation procedures
for the particular case of the S-wave two-nucleon contact-range
interactions.
Standard saturation is known to work well when comparing the EFT
contact-range couplings with a series of
OBE potentials~\cite{Epelbaum:2001fm}.
Thus the natural question at this point is whether this is still
the case with the novel saturation method of Ref.~\cite{Peng:2020xrf}.

However nucleons are not as heavy as the charmed mesons and baryons,
which means that relativistic corrections are often included
in the light-meson exchange potentials.
This implies that the saturation formulas we have to use are somewhat
more involved than the ones we obtained in Sect.~\ref{sec:pc}.
Actually, the corresponding formulas for the standard saturation method
(which also include form factors) can be found
in Ref.~\cite{Epelbaum:2001fm}.
It happens that the novel saturation method merely generates and additive
factor in the standard saturation relations of Ref.~\cite{Epelbaum:2001fm}:
\begin{eqnarray}
  C^{\rm sat(novel)} = C^{\rm sat(standard)} + \delta C^{\rm sat} \, .
  \label{eq:C-additive}
\end{eqnarray}
Here we notice that, for potentials of the ${\vec{q}^2} / (m^2 +\vec{q}^2)$
type, the novel saturation procedure can be encapsulated
in the following substitution rule
\begin{eqnarray}
  \frac{\vec{q}^2}{\vec{q}^2 + m^2} \to \frac{\vec{q}^2}{\vec{q}^2 + m^2} - 1
  \, ,
  \label{eq:V-subs}
\end{eqnarray}
which merely amounts to the inclusion of an additive term in the potential
to manually remove the Dirac-delta term, thus justifying
the rule in Eq.~(\ref{eq:C-additive}).
Now in the two-nucleon system we also find relativistic corrections
that follow the general form
\begin{eqnarray}
  \frac{\vec{k}^2}{\vec{q}^2 + m^2} = - \frac{1}{4}\,
  \frac{\vec{q}^2}{\vec{q}^2 + m^2} +
  \frac{1}{4}\,\frac{\vec{p}^2 + \vec{p}'^2}{\vec{q}^2 + m^2} \, ,
\end{eqnarray}
which we have rewritten as the sum of a purely local and non-local terms
(i.e. the terms proportional to $\vec{q}^2$ and $(\vec{p}^2+\vec{p}'^2)$,
respectively).
For the saturation of the local term we will use again
the substitution rule of Eq.~(\ref{eq:V-subs}).

From the previous, the modifications for a scalar, pseudoscalar and
vector meson are
\begin{eqnarray}
  \delta V_S &=& \frac{g_S^2}{4 M_N^2} F^2_S(\vec{q}^2) \, , \\
  \delta V_P &=& \frac{g_P^2}{12 M_N^2} \vec{\sigma}_1 \cdot \vec{\sigma}_2 \,
  F^2_P(\vec{q}^2) \, , \\
  \delta V_V &=& \Big[ \frac{g_V (g_V + f_V)}{2 M_N^2} + \frac{2}{3}
    {\left( \frac{g_V + f_V}{2 M_N}\right)}^2 \vec{\sigma}_1 \cdot \vec{\sigma}_2
    \Big]
  F^2_V(\vec{q}^2) \, , \nonumber \\
\end{eqnarray}
where $M_N$ is the nucleon mass, $g_S$ ($g_P$) the coupling constant for
the scalar (pseudoscalar) meson, $g_V$ and $f_V$ the electric- and
magnetic-like couplings for the vector meson, $\vec{\sigma}_{1(2)}$
the spin operators for nucleons $1(2)$ and $F_M(\vec{q}^2)$
the form-factors, which accept the expansion
\begin{eqnarray}
  F_M(\vec{q}^2) = \alpha_1 + \alpha_2 \frac{\vec{q}^2}{\Lambda_M^2} + \dots
\end{eqnarray}
with $\Lambda_M$ the form-factor cutoff for the particular meson $M$
under consideration.
If the exchanged meson is an isovector (e.g. the $\rho$),
the isospin operator $\vec{\tau}_1 \cdot \vec{\tau}_2$
will have to be included in the previous expressions.

If we consider the momentum expansion of the S-wave contact-range potential
\begin{eqnarray}
  \langle p' | V_C | p \rangle = C_0 + C_2\,(p'^2 + p^2) + \dots \, ,
\end{eqnarray}
the novel saturation procedure generates the following modifications
for the $^1S_0$ and $^3S_1$ partial waves in the two-nucleon system
(where we have used the spectroscopic notation ${}^{2S+1}L_J$ with
$S$, $L$, $J$ the intrinsic, orbital and total spin,
respectively).
For a scalar meson, we have to add the terms
\begin{eqnarray}
  \delta C_{0S}^{{}^1S_0} &=&  \frac{g_S^2}{4 M_N^2} \alpha_1^2 \, , \\
  \delta C_{2S}^{{}^1S_0} &=&  \frac{g_S^2}{2 M_N^2}
  \frac{\alpha_1 \alpha_2}{\Lambda_S^2} \, , \\
  \nonumber \\
  \delta C_{0S}^{{}^3S_1} &=&  \delta C_{0S}^{{}^1S_0} \, , \\  
  \delta C_{2S}^{{}^3S_1} &=&  \delta C_{2S}^{{}^1S_0} \, ,
\end{eqnarray}
while for a pseudoscalar meson we will have
\begin{eqnarray}
  \delta C_{0P}^{{}^1S_0} &=&  -\frac{g_P^2}{4 M_N^2} \alpha_1^2 \, , \\
  \delta C_{2P}^{{}^1S_0} &=&  -\frac{g_P^2}{2 M_N^2}
  \frac{\alpha_1 \alpha_2}{\Lambda_P^2} \, , \\
  \nonumber \\
  \delta C_{0P}^{{}^3S_1} &=&  -\frac{1}{3}\delta C_{0P}^{{}^1S_0} \, , \\  
  \delta C_{2P}^{{}^3S_1} &=&  -\frac{1}{3}\delta C_{2P}^{{}^1S_0} \, .
\end{eqnarray}
For vector meson exchange we add the terms
\begin{eqnarray}
  \delta C_{0P}^{{}^1S_0} &=& -\frac{f_V (g_V + f_V)}{2 M_N^2}\,\alpha_1^2 \, , \\
  \delta C_{2P}^{{}^1S_0} &=& 
  -\frac{f_V (g_V + f_V)}{M_N^2}\,
  \frac{\alpha_1 \alpha_2}{\Lambda_V^2} \, , \\
  \nonumber \\
  \delta C_{0P}^{{}^3S_1} &=& \frac{1}{6\,M_N^2}\,
  \left( 4 g_V^2 + 5 g_V f_V + f_V^2\right)\,\alpha_1^2 \, , \\
  \delta C_{2P}^{{}^3S_1} &=&
  \frac{1}{3\,M_N^2}\,
  \left( 4 g_V^2 + 5 g_V f_V + f_V^2\right)\,
  \frac{\alpha_1 \alpha_2}{\Lambda_V^2} \, .
\end{eqnarray}
Finally, Ref.~\cite{Epelbaum:2001fm} uses the following definition
for the coupling constants
\begin{eqnarray}
  \hat{C} = 4\pi \, C_0 \quad \mbox{and} \quad C = 4 \pi \, C_2 \, ,
\end{eqnarray}
which we will also use in what follows for a more convenient comparison.

Putting all the pieces together, for the particular case of
the Bonn B potential~\cite{Machleidt:1987hj} we obtain
\begin{eqnarray}
  \hat{C}_{{}^1S_0}^{\rm novel} &=& \hat{C}_{{}^1S_0}^{\rm std} + \delta \hat{C}_{{}^1S_0}
  \nonumber \\
  &=& (- 0.117 - 0.134 ) \cdot 10^{4}\,{\rm GeV}^{-2}  \nonumber \\
  &=& -0.251 \cdot 10^{4}\,{\rm  GeV}^{-2} \, , \\
  \nonumber \\
  {C}_{{}^1S_0}^{\rm novel} &=& {C}_{{}^1S_0}^{\rm std} + \delta {C}_{{}^1S_0}
  \nonumber \\
  &=& (1.276 + 0.178 ) \cdot 10^{4}\,{\rm GeV}^{-4}  \nonumber \\
  &=& 1.454 \cdot 10^{4}\, {\rm  GeV}^{-4} \, , 
\end{eqnarray}
for the singlet, to be compared with
$\hat{C}_{{}^1S_0}^{\rm N^2LO} = \{ -0.160, -0.158 \,\cdot 10^{4}\,{\rm GeV}^{-2}$
and ${C}_{{}^1S_0}^{\rm N^2LO} = \{ 1.134, 1.135 \} \cdot 10^{4}\,{\rm  GeV}^{-4}$
(i.e. the results for the EFT couplings in the two-nucleon system at
next-to-next-to-leading ($\rm N^2LO$) as determined
in Ref.~\cite{Epelbaum:2001fm}, where the brackets indicate
their expected variation within the formalism of the
aforementioned reference), while for the triplet we get
\begin{eqnarray}
  \hat{C}_{{}^3S_1}^{\rm novel} &=& \hat{C}_{{}^1S_0}^{\rm std} + \delta \hat{C}_{{}^1S_0}
  \nonumber \\
  &=& (- 0.101 - 0.091 ) \cdot 10^{4}\,{\rm GeV}^{-2}  \nonumber \\
  &=& -0.192 \cdot 10^{4}\,{\rm  GeV}^{-2} \, , \\
  \nonumber \\
  {C}_{{}^3S_1}^{\rm novel} &=& {C}_{{}^1S_0}^{\rm std} + \delta {C}_{{}^1S_0}
  \nonumber \\
  &=& (0.660  + 0.178) \cdot 10^{4}\,{\rm GeV}^{-4}  \nonumber \\
  &=& 0.838 \cdot 10^{4}\,{\rm  GeV}^{-4} \, , 
\end{eqnarray}
to be compared with
$\hat{C}_{{}^3S_1}^{\rm N^2LO} = \{ -0.159, -0.134 \} \cdot 10^{4}\,{\rm  GeV}^{-2}$
and
${C}_{{}^3S_1}^{\rm N^2LO} = \{ 0.637, 0.587 \} \cdot 10^{4}\,{\rm  GeV}^{-4}$
for EFT.
From this, in the case of the singlet channel with the Bonn-B potential,
the novel saturation method underperforms the standard one,
while for the triplet channel their deviations
with respect the EFT couplings are similar.
However this comparison is potential-dependent:
Ref.~\cite{Epelbaum:2001fm} considers a total of six phenomenological
potentials, while here we limit ourselves to Bonn-B,
i.e. the easiest one on which to apply saturation.


%

\end{document}